%% file: sample-sigconf.tex
\documentclass[sigconf,natbib=true,anonymous=false]{acmart}
\AtBeginDocument{%
  }

\copyrightyear{2026}
\acmYear{2026}
\setcopyright{cc}
\setcctype{by}
\acmConference[SIGSPATIAL '26]{The 34th ACM International Conference on Advances in Geographic Information Systems}{November 03--06, 2026}{Riverside, CA, USA}
\acmBooktitle{The 34th ACM International Conference on Advances in Geographic Information Systems (SIGSPATIAL '26), November 03--06, 2026, Riverside, CA, USA}
\acmDOI{10.1145/3841645.3844187}
\acmISBN{979-8-4007-2950-8/2026/11}

\usepackage{amsmath}
\usepackage{subfigure}
\usepackage[dvipsnames]{xcolor}
\usepackage{fontawesome}
\usepackage{placeins} 
\usepackage{ulem}
\usepackage{microtype}
\usepackage{graphbox}
\usepackage{xspace}
\usepackage[most]{tcolorbox}
\usepackage{tikz}
\usepackage{pifont}
\usepackage{multirow}
\usepackage{tikz}
\usetikzlibrary{patterns}
\usepackage{nicematrix}
\usepackage{algorithm}
\usepackage{algpseudocode}
\usepackage{tabularx}
\usepackage{booktabs} 
\usepackage{multirow} 
\usepackage{float}    
\usetikzlibrary{patterns} 
\tcbuselibrary{skins} 

\newcommand{\best}[1]{\underline{\textbf{#1}}}

\newcommand{\quotePrompt}[1]{{\tiny\faQuoteLeft}\textit{#1}{\scriptsize\textsuperscript{\faQuoteRight}}}

\newcommand{\partitle}[1]{\paragraph{\hspace{-11pt}\textbf{#1}}}

\newcommand{\parnone}{\par\medskip\hspace{-2pt}}

\definecolor{ptBlue}{RGB}{0.117, 0.564, 1.0}
\definecolor{ptOrange}{RGB}{1.0, 0.55, 0.0}
\definecolor{ptGreen}{RGB}{0.133, 0.545, 0.133}
\definecolor{redbrown}{RGB}{165, 42, 42}

\newcommand{\xmark}{\textcolor{redbrown}{\ding{55}}}

\newcommand{\boxpromptnumbered}[2]{%
  \begin{tcolorbox}[
    enhanced,
    colback=white,
    colframe=black!70,
    boxrule=1.2pt,
    arc=3pt,
    left=6pt, right=1pt, top=6pt, bottom=1pt,
    overlay={
      \node[
        rectangle,           
        fill=white,          
        text=black!70,       
        font=\bfseries\Small,
        inner sep=2pt        
      ] at (frame.north west) {#1};
    }
  ]
  \Small
  \texttt{#2}
  \end{tcolorbox}%
}

\newcommand{\opt}[1]{\textcolor{black!50}{#1}}

\begin{document}

\title{Do We Still Need Gazetteers in the Era of LLMs? Chaining Retrieval with a Spatial Neuro-Symbolic Index}

\input{00_CONTENT/01_Authors}

\input{00_CONTENT/02_Abstract}

\input{00_CONTENT/03_Metadata}


\begin{teaserfigure}
    \centering
    \includegraphics[width=0.9\columnwidth]{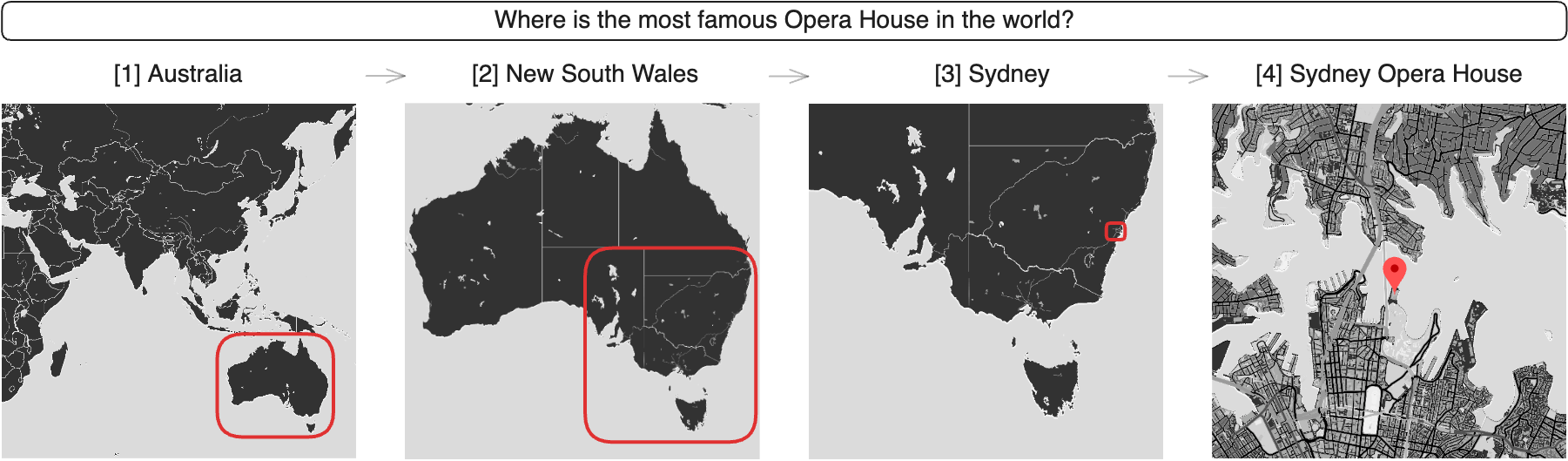}
    \caption{
    In search systems, an index structures data to map query keys to candidate records. Traditionally, a lexical index relies on surface string matching (e.g., resolving ``Opera House'' to ``Royal Opera House'' in London), whereas a spatial index tolerates candidates across varying geographic resolutions (e.g., validating the broader region of ``New South Wales''). In this paper, a spatial-semantic index merges both modalities to retrieve conceptually \textit{accurate} candidates across all hierarchical scales with varying spatial \textit{precision}. This concept of an index generalizes beyond traditional spatial database indexing (e.g., geometry-based structures) \cite{sametFoundationsMultidimensionalMetric2005} to support effective Geographic Information Retrieval (GeoIR) in text-based applications.
    }
    \Description{Indexing}
    \label{fig:example_opera_house}
\end{teaserfigure}

\maketitle

\input{00_CONTENT/04_Intro}

\input{00_CONTENT/05_Related_works}
\input{00_CONTENT/06_Problem_statement}
\input{00_CONTENT/07_Methodology}
\input{00_CONTENT/08_Results}

\input{00_CONTENT/09_Discussions}

\input{00_CONTENT/10_Limitations}

\input{00_CONTENT/11_Conclusion}
\begin{acks}
This research was undertaken with the assistance of computing resources from RACE (RMIT Advanced Cloud Ecosystem). Gemini Pro was used for code generation and text polishing. We thank the four anonymous reviewers for their time and valuable insights.
\end{acks}

\clearpage

\bibliographystyle{ACM-Reference-Format}
\bibliography{sample-base}

\appendix

\input{00_CONTENT/A_Appendix}

\end{document}

%% file: 00_CONTENT/01_Authors.tex
\author{Alexis Horde-Vo}
\orcid{0009-0004-3324-3380}
\affiliation{%
  \institution{RMIT University}
  \city{Melbourne}
  \country{Australia}
}
\email{alexis.horde.vo@student.rmit.edu.au}

\author{Matt Duckham}
\orcid{0000-0002-7249-6709}
\affiliation{%
  \institution{RMIT University}
  \city{Melbourne}
  \country{Australia}
}
\email{matt.duckham@rmit.edu.au}

\author{Estrid He}
\orcid{0000-0002-8994-9532}
\affiliation{%
  \institution{RMIT University}
  \city{Melbourne}
  \country{Australia}}
\email{estrid.he@rmit.edu.au}


%% file: 00_CONTENT/02_Abstract.tex


\begin{abstract}
Geographic information retrieval (GeoIR) tasks require systems to interpret ambiguous toponyms for downstream applications.
Traditionally, toponym resolution relies on gazetteers to provide an explicit index of place entities and spatial relationships. 
Recently, gazetteer-free approaches seek to reduce dependence on handcrafted searches: dense retrieval utilizes text encoders to capture rich context, moving beyond the limitations of lexical search.
However, text encoders implicitly assume that learned representations can function as reliable spatial-semantic indexes.
In this paper, we evaluate this assumption through a spatial-semantic indexing setup: given a contextualized toponym mention, we retrieve the corresponding gazetteer entity represented by text derived from a gazetteer knowledge graph. 
We benchmark five frozen text encoders under two retrieval strategies: brute-force nearest-neighbor retrieval over entity representations, and a neuro-symbolic hierarchical beam search that constrains retrieval (i.e. chaining the search with gazetteer hierarchy).
Experimental results reveal a distinct coarse-versus-fine trade-off. Unconstrained dense retrieval frequently incurs catastrophic spatial errors. Conversely, hierarchical constraints improve coarse geographic grounding, but still yield limited benefit for fine-grained localization metrics: vanilla text encoders fail to capture the fine-scale spatial fidelity encoded in gazetteers.
Our code is publicly available at: 
\url{https://doi.org/10.25439/rmt.31094269}

\end{abstract}

%% file: 00_CONTENT/03_Metadata.tex
\ccsdesc[500]{Information systems~Information retrieval}
\ccsdesc[300]{Information systems~Geographic information systems}

\keywords{Multi-vector retrieval, semantic indexing, toponym, geographic information retrieval, gazetteer.}



%% file: 00_CONTENT/04_Intro.tex
\section{Introduction}


Gazetteers link place names (toponyms) to geographic entities and their locations. These structured indexes underpin many core GeoIR tasks, including toponym resolution and geocoding. 
However, resolving a toponym from text is 
difficult: 
names exhibit spatial ambiguity with underspecification (e.g., ``Sydney'' in Australia or Canada?),
and the surface form in text may not exactly match gazetteer entries.
Alternatively, recent pretrained text encoders provide rich contextual and semantic representations,
and act as learned indexes \cite{kraskaCaseLearnedIndex2018, jinLanguageModelsSemantic2024}. Indeed, these models map toponyms to continuous vector spaces, thereby minimizing the search operations required for toponym resolution.


Currently, retrieving spatial information from text relies on two paradigms. First, \textit{knowledge-based question answering} (KB-QA) uses strict symbolic mapping with a machine-executable form (e.g., SQL, SPARQL, and Cypher, notably in spatial-RAG applications \cite{yuSpatialRAGSpatialRetrieval2025}), but is brittle against underspecified queries \cite{shermanGenerativeAIGeospatial2025}. Second, \textit{information-retrieval-based question answering} (IR-QA) employs dense similarity search over learned representations \cite{lewisRetrievalAugmentedGenerationKnowledgeIntensive2021, khattabColBERTEfficientEffective2020, karpukhinDensePassageRetrieval2020b, scheererWARPEfficientEngine2025}.
For spatial tasks, dense retrieval is more generalizable:
models rely on distributional semantics and contextual cues without requiring an exact symbolic match. Nonetheless, similarity-based retrieval does not inherently enforce geographic constraints: \textbf{semantically plausible candidates can be geographically implausible}, and small representation errors can lead to large spatial errors (Table~\ref{tab:geo_results}).

\input{00_CONTENT/FIGS_TABLES_IMG_ALG/tab_ranking_australia}

This limitation is critical in modern multi-objective pipelines, such as open-domain question answering \cite{voorheesTRECQuestionAnswering2001} or recommendation \cite{tangUGuideRAGIntentEnhancedRetrievalAugmented2025}. 
Solutions rely on lightweight text encoders for rapid initial lookups \cite{khattabColBERTEfficientEffective2020, scheererWARPEfficientEngine2025} first; then followed by resource-intensive LLMs for complex reasoning \cite{weiChainofThoughtPromptingElicits2023, asaiSelfRAGLearningRetrieve2023a};
and toponym resolution only serves as an ancillary signal throughout the process.
These dependencies create a severe performance bottleneck: the entire pipeline fails if the primary text encoders omit relevant geographic cues \cite{kandpalLargeLanguageModels}.

The study evaluates text encoders as a core spatial indexing module and raises the key question: \textbf{to what extent can text encoders support accurate spatial-semantic indexing of places without relying on explicit spatial structure (e.g., a gazetteer)?}

In this paper, we 
isolate the problem 
through the lens of \textit{spatial-semantic indexing} \cite{kritharaRoadManualAutomatic2023} (Figure~\ref{fig:example_opera_house}): given a spatial expression (a toponym mention in context), 
we retrieve the corresponding geographic entity from a gazetteer organized as a knowledge graph.
We consider two retrieval settings. First, we employ \textit{an unconstrained baseline} that performs brute-force nearest-neighbor search over all entity representations. Second, we introduce \textit{a neuro-symbolic retrieval strategy} that uses hierarchical constraints from the gazetteer to structure the search via hierarchical beam search, reducing confusion among geographically incompatible candidates while retaining the encoder's ability to interpret context.

The concept of error in GeoIR diverges from traditional Information Retrieval (IR) \cite{duckhamFormalOntologicalApproach2000, maiGeographicQuestionAnswering2021a}: \textit{accurate} toponym matching identifies a specific database entity and, by extension, all dependent entities; whereas \textit{precise} matching targets any object within a distance threshold.
In contrast to existing work, we deliberately enforce a cut-off resolution (pruning) to prioritize conceptual alignment over simple string or coordinate matching, and measure true conceptual generalization over memorization of the reference dataset.

Across \textit{five} text encoders and \textit{four} benchmark datasets, we show that unconstrained embedding retrieval is often insufficient for reliable toponym resolution, notably for spatial-semantic indexing.
In contrast, incorporating explicit spatial structure through gazetteer hierarchies improves accuracy. 
More broadly, the findings support a complementary view: \textit{text encoders contribute semantic generalization for interpreting spatial mentions, while gazetteer structure provides geographic fidelity and principled constraints during retrieval}. Gazetteers are indispensable for spatial grounding rather than being replaced by language models.

\partitle{Contributions.}
The paper makes three main contributions: 
\begin{enumerate}
    \item the construction of an extensible neuro-symbolic index of toponyms for spatial-semantic indexing tasks, structured as a knowledge graph modeling containment, direction, and textual relations;
    \item an empirical evaluation of language models, to quantify how gazetteer structure improves alignment between geographic space and the language model embedding space, by chaining dense retrieval; and
    \item an assessment of toponym resolution generalization in semantic indexing, measured by pruning the target gazetteer.
\end{enumerate}

%% file: 00_CONTENT/FIGS_TABLES_IMG_ALG/tab_ranking_australia.tex
\begin{table}[bt]
    \centering
    \caption{The encoder EmbeddingGemma-300M prioritizes exact matches (``Australia'') and orthographically similar terms (``Austria''), but fails to retrieve relevant sub-country levels such as the Northern Territory, New South Wales, or Tasmania. In addition, the artifact ``Nelson'' is better ranked than any major Australian city.}
    \label{tab:geo_results}
    {\small
    \begin{tabular}{@{} p{0.48\linewidth} p{0.48\linewidth} @{}}
        \toprule
        \multicolumn{2}{@{} p{\linewidth} @{}}{\textbf{Query:} \quotePrompt{I want to find geographic descriptions about Australia.}} \\ 
        \multicolumn{2}{@{} p{\linewidth} @{}}{\textbf{Document 5:} \quotePrompt{\underline{\textbf{Victoria}} is located in \underline{Australia}. \underline{\textbf{Victoria}} is surrounded by \underline{South Australia, etc}. Coasts of \underline{\textbf{Victoria}} include \underline{Bass Strait}.}} \\
        \midrule
        \multicolumn{2}{@{} p{\linewidth} @{}}{\textbf{Top-10 place candidates are related to the documents describing:}} \vspace{1mm} \\ 
        (1)~Australia & (6)~South Australia \\
        (2)~Australian Capital Territory & (7)~Queensland \\
        (3)~Great Australian Bight & (8)~Nelson (New Zealand) {\color{BrickRed}\xmark} \\
        (4)~Western Australia & (9)~Austria {\color{BrickRed}\xmark} \\
        (5)~Victoria & (10)~South Pacific Ocean {\color{BrickRed}\xmark} \\
        \bottomrule
    \end{tabular}
    }
\end{table}

%% file: 00_CONTENT/05_Related_works.tex
\section{Related work}



Any place name is ambiguous. The dependency on both footprint extent and topical context has challenged the creation of integrated spatial and \textit{textual} indexing methods for decades \cite{jonesGeographicalInformationRetrieval2008a}. 

Related work in sensitive domains  enhances technical text understanding with semantic indexing (e.g., the biomedical field \cite{kritharaRoadManualAutomatic2023,tsatsaronisOverviewBIOASQLargescale2015}): 
disambiguation maps an expression in context to external, unequivocal knowledge bases of reference, and ideally those curated with ontologies \cite{chandrasekaranWhatAreOntologies1999}. 
Spatial interpretation is also particularly sensitive. Indeed, divergences across different individuals' spatial cognition \cite{tverskyCognitiveMapsCognitive1993} or social conceptions \cite{doussetSystemsGeographyGeography2018} are all potential sources of misinterpretation that impact retrieval quality from the perspective of spatial-semantic alignment. 

In GeoIR applications, gazetteers \cite{goodchildIntroductionDigitalGazetteer2008} define the geographic reference:
gazetteers index the vocabulary of toponyms, by integrating resources from local authorities, volunteered geographic information \cite{kesslerAgendaNextGeneration2009}, or historical records \cite{grossnerLinkedPlacesWorld2021, rusinekFeedingGazetteerLeveraging2021}. 
Semantic indexing task can help to standardize toponym understanding in text as a toponym resolution task: linking a spatial mention to a gazetteer entry \cite {liebermanGeotaggingLocalLexicons2010}. To this end, 
two approaches are used for toponym resolution: 
knowledge-base question answering (KB-QA) and information-retrieval question-answering (IR-QA).

\partitle{KB-QA}
Traditional systems rely on handcrafted heuristics to query gazetteers as structured place databases \cite{zhangSurveyGeocodingAlgorithms2025,vaidSpatiotextualIndexingGeographical2005}. 

Current methods operate primarily as \textit{filters} within the KB-QA paradigm, relying on sophisticated pattern matching models at early stages, and focusing on homonymy resolution in a second stage. In the SemEval 2019 task for \textit{Toponym Resolution in Scientific Papers}, the best systems first pre-filtered candidates from the GeoNames database \cite{wangDM_NLPSemEval2018Task2019, liUniMelbSemEval2019Task2019, yadavUniversityArizonaSemEval20192019a}.
Purely lexical matching fails if the mention does not closely match any canonical entry due to lexical variation or underspecification \cite{catfordLinguisticTheoryTranslation,janssenMULTILINGUALLEXICALDATABASES2004}. Practical pipelines mitigate lexical mismatches by optimizing queries through fuzzy string matching \cite{zhangImprovingToponymResolution2023} or place normalization \cite{linRACCOONRetrievalAugmentedGeneration2025} as preliminary steps, before employing BM25-style keyword retrieval \cite{linRACCOONRetrievalAugmentedGeneration2025} or sophisticated rule-based geocoders \cite{huToponymResolutionLeveraging2024}. 

Such optimizations improve matching performance but remain overly tailored to specific toponym resolution tasks \cite{wangAreWeThere2020}; to applications requiring exact locations (e.g., location-based services); and to the quality of the chosen gazetteer \cite{achesonQuantitativeAnalysisGlobal2017}. For extensive retrieval, these methods incur prohibitive computational costs \textit{at scale} \cite{huToponymResolutionLeveraging2024}, or operate as binary filters rather than \textit{rankers}.


\partitle{IR-QA}
Early work on latent semantic analysis (LSA) \cite{deerwesterIndexingLatentSemantic1990} and subsequent neural representation learning aimed to capture synonymy beyond exact term overlap. Modern transformer encoders \cite{vaswaniAttentionAllYou2017} provide contextual representations to enable dense (and, in some cases, multi-vector) retrieval over candidate entities.
Consequently, the widespread adoption of text embedding models as retrieval encoders \cite{chenBGEM3EmbeddingMultiLingual2024a,veraEmbeddingGemmaPowerfulLightweight2025,zhangQwen3EmbeddingAdvancing2025,behnamghaderLLM2VecLargeLanguage2024} 
positions learned representations as ``approximate learned'' indexes for retrieval \cite{jinLanguageModelsSemantic2024,al-mamunSurveyLearnedIndexes2024}.

In GeoIR, encoder-based retrieval systems
utilize end-to-end models to embed geographical structure within latent representations, 
disambiguate toponyms, and generalize across lexical variants \cite{gocmenAdvancingSpatialKeyword}. 
Hence, IR-QA methods are gazetteer-free during inference.

Early methods can even bypass traditional gazetteers and map text directly to geometric indexes, e.g. discretized global grid cells.
For example, CamCoder \cite{grittaWhichMelbourneAugmenting2018}
and multi-level geocoding (MLG) models \cite{kulkarniMultiLevelGazetteerFreeGeocoding2021}
utilize text representations (respectively $n$-grams and GloVe embeddings \cite{penningtonGloVeGlobalVectors2014}) and CNN classifiers to activate spatial cells.
These approaches reduce toponym resolution to a coordinate retrieval task \cite{grittaPragmaticGuideGeoparsing2020}.
However, geocoding focuses on only one spatial property (i.e., geometry) and does not fully resolve spatial ambiguities. For example, a single pair of coordinates can represent multiple spatial objects at different resolutions or precisions, neglecting gazetteer semantics.
More recently, contrastive fine-tuning of sentence encoders improves toponym resolution at
fine-grained \cite{nakataniTextEmbeddingModel} and global \cite{gomesFineTuningTransformersToponym} levels;
however, evaluations rely on narrow geographic setups: the spatial scope is limited to a single country \cite{nakataniTextEmbeddingModel}; or methodological choices artificially inflate model confidence by partitioning highly specialized datasets (with a specific set of homonymous toponyms) and testing with limited variability compared to the training dataset.

\medskip
Finally, current research often treats toponym resolution as a standard entity linking task, where exact string matching conflates data memorization with true spatial generalization.  Evaluations face two main gaps: (1) the capability of models to resolve entities across complex gazetteer \textit{hierarchies} is not assessed as an essential requirement for spatial-semantic indexing; and (2) the narrow geographic scope of most benchmark datasets prevents accurate assessment of global \textit{generalization}. 
Modeling gazetteers as knowledge graphs allows us to validate model calibration across diverse graph traversal strategies. In this setup, multi-level graph sampling serves as a data augmentation mechanism, evaluating whether models preserve geographic fidelity while maintaining robust spatial generalization.

%% file: 00_CONTENT/06_Problem_statement.tex
\section{Task Definition}
Let $T$ be an input text (e.g., a sentence or paragraph) that contains a spatial expression $t$ (a toponym mention such as a place name). The goal of \textbf{spatial-semantic indexing} is to resolve $t$ in the context $T$ and link it to the most suitable geographic entity in a reference inventory of places (e.g., a gazetteer). 
Let $P$ be a finite set of candidate geographic entities from a gazetteer available for linking. The task can be formulated as:
\begin{equation}
    \label{def:problem}
    \hat{p} = \arg\max_{p \in P} \; \texttt{Rel}\big([t,T], p\big)
\end{equation}
where $\texttt{Rel}(\cdot)$ denotes a relevance function that scores how well candidate entity $p$ corresponds to the mention $t$ as used in text $T$. 
Unlike standard toponym resolution, spatial-semantic indexing tolerates gazetteer pruning by falling back to the most relevant parent entity when an exact match is absent.

In our setting, the candidate set $P$ is embedded
in a geographic knowledge graph $\mathcal{K}=(\mathcal{V}, \mathcal{E}, \mathcal{R})$, where $\mathcal{V}$ is the set of nodes (including geographic entities, i.e., $P \subseteq \mathcal{V}$), $\mathcal{R}$ is a set of relation types (e.g., \texttt{locatedIn}, \texttt{partOf}, \texttt{adjacentTo}, \texttt{northOf}, \texttt{hasDescription}, and \texttt{hasGeometry} with H3 cells), and $\mathcal{E} \subseteq \mathcal{V} \times \mathcal{R} \times \mathcal{V}$ is the set of labeled edges. 

%% file: 00_CONTENT/07_Methodology.tex
\section{Methodology}


Equation~\ref{def:problem} requires defining a toponym $t$ to be disambiguated, candidate geographic entities $p$, and a relevance function.
Section~\ref{section:benchmark_dataset} presents the benchmark datasets for toponym resolution with toponyms $t$ in context.
Section~\ref{section:kg_creation} enriches each toponym with a textual description $p$ for matching: first from gazetteer extraction (geospatial view), and second from Wikipedia (Wikipedia view). 
Section~\ref{section:methodology_lm_retrieval} defines two relevance functions, based on brute-force search and hierarchical beam search (HBS).
Section~\ref{section:eval} outlines the metrics for evaluation.

\subsection{Benchmark datasets}
\label{section:benchmark_dataset}

We use four benchmark datasets\footnote{Provided by the platform EUPEG \cite{wangEnhancingSpatialTextual2019b} containing unstructured texts ($T$) with annotated spatial mentions ($t$), where each mention is linked to a ground-truth point location (latitude, longitude): \hyperlink{https://github.com/geoai-lab/EUPEG}{https://github.com/geoai-lab/EUPEG}}: 
\begin{itemize}
    \item GeoVirus (\textbf{GV}) \cite{grittaWhichMelbourneAugmenting2018}: 229 WikiNews articles in 2017 related to disease and epidemics with 2k 
    manual annotations;
    \item Lexical Global Lexicon (\textbf{LGL}) \cite{liebermanGeotaggingLocalLexicons2010}: 588 articles from 78 newspapers with highly ambiguous toponyms (homonyms), with 5k manual annotations  
     (1k homonyms)
    whose 41\% of the toponyms represent small populated places at a local level;
    \item TR-News (\textbf{TRN}) \cite{kamalooCoherentUnsupervised2018}: 118 documents from 36 sources containing 
    1k annotated toponyms (352 homonyms);
    \item WikTor (\textbf{WT}) \cite{grittaWhatsMissingGeographical2018}: 5000 automatically extracted Wikipedia documents describing 5000 toponyms selected for their ambiguity (1906 unique homonyms), with 25k total annotations.
\end{itemize}

We created a context $T$ (window: 50 words) for each toponym $t$:  
\boxpromptnumbered{}{Let's focus on the location **\{\textit{t}\}** in the text: ``\{\textit{T}\}''. Context: **\{\textit{t}\}**.} 

These datasets contain contemporary mentions for toponyms. 

\subsection{Textual toponym enrichment}
\label{section:kg_creation}

\begin{figure}[hbt!]
    \centering
    \includegraphics[width=\columnwidth]{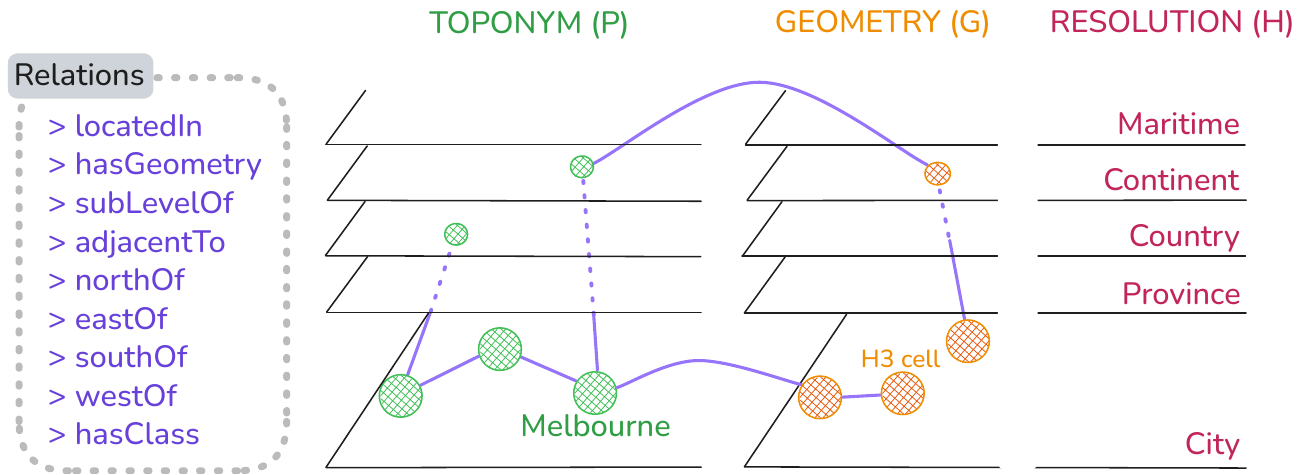}
    \caption{Representation of the ontology of $\mathcal{K}$.}
    \Description{Ontology}
    \label{fig:KG_onto}
\end{figure}

We assume a reference geographic knowledge graph $\mathcal{K}=(\mathcal{V},\mathcal{E},\mathcal{R})$ that organizes candidate toponyms and their spatial relations. As illustrated in Figure~\ref{fig:KG_onto}, nodes in $\mathcal{K}$ include: (i) toponym entities $p \in P$, (ii) geometry nodes $g \in G$ that represent the spatial footprint of entities at a chosen granularity, and (iii) a small set of hierarchy/level nodes $h \in H$ that encode coarse semantic categories or administrative levels (e.g., \texttt{Maritime}, \texttt{Continent}, \texttt{Country}, \texttt{Province}, and \texttt{City}). Thus, $\mathcal{V} = P \cup G \cup H$.

Edges $\mathcal{E}$ use relation types $\mathcal{R}$ to capture both hierarchical and spatial structure. In particular, we model:
\textbf{Entity-to-geometry links} (e.g., \texttt{locatedIn}/\texttt{hasGeometry}) connecting a toponym $p$ to H3 cells $g$; \textbf{Hierarchical containment} between toponyms (e.g., \texttt{subLevelOf}), enabling traversal from coarse to fine geographic scopes; \textbf{Spatial neighborhood relations} between geometry nodes (e.g., \texttt{adjacentTo}, directional relations (e.g., \texttt{northOf}); and \textbf{Type/level metadata} (e.g., \texttt{hasClass} linking to a level node in $H$) to support level-aware retrieval.

\partitle{Source data.} We instantiated $\mathcal{K}$ from two public gazetteer sources: Global Administrative Areas (GADM)\footnote{\url{https://gadm.org/}} and the IHO World Seas dataset from MarineRegions\footnote{\url{https://www.marineregions.org/downloads.php}}. From GADM, we retained administrative units up to level 2, and we merged these with IHO maritime regions to form the candidate toponym inventory. Each retained entry is normalized to a canonical name (and aliases when available) and assigned a class\footnote{Note that the classification by GADM might be locally inaccurate (e.g., overseas territories can be considered as countries, or the city level may correspond to wider urban areas two administrative levels below the country level). For example, in France, the city level is actually a \textit{department} while in Australia it is a \textit{local government area}. This problem is due to the absence of a richer ontology in the datasets \cite{janowiczRoleOntologyImproving2008}. However, we retain the labels in $H$ for readability, as these inaccurate labels do not affect the hierarchy.}
label in $H$ = \{\textit{Maritime}, \textit{Continent}, \textit{Country}, \textit{Province}, \textit{City}\} for level-aware retrieval.

To reduce reliance on exact coordinate matching and to standardize spatial footprints across sources, we discretized gazetteer point geometries into H3 cells up to the resolution 3 (city-level granularity). We then derived geometry adjacency and coarse directional relations directly from the H3 grid. 

The resultant knowledge graph contains 51,249 toponyms in $P$ (101 maritime areas, 7 continents, 263 countries, 3,661 provinces, and 47,217 cities), and 30,067 H3 cells in $G$. 

To use text encoders for retrieval, we associated each candidate toponym $p \in P$ with a textual description 
$D_{p, \textrm{geo}}$ via GeoView or $D_{p, \textrm{wiki}}$ via WikiView.

\partitle{GeoView} 
The semantic distribution of toponyms is uneven in human writing, with an emphasis on popular places.
For example, \textit{Sydney} is more frequently written alongside the distant \textit{Melbourne} rather than the suburb \textit{Canada Bay}. For each toponym, the geospatial view (GeoView) creates text documents that contain a neutral and unbiased description based solely on gazetteers.
Hence, we derive a document $D_{p, \textrm{geo}}$ deterministically from the KG neighborhood of the toponym $p$. Intuitively, $D_{p, \textrm{geo}}$ summarizes (a) the entity name, (b) its hierarchical path (parents/containers), and (c) salient spatial relations (adjacency and/or directional relations) expressed as short natural language using a fixed template:
\boxpromptnumbered{GEO}{location: Australia / sub-administrative level of: Oceania / adjacent to: Indian Ocean\dots / to the north: Coral Sea\dots}


\partitle{WikiView} Toponyms are also lexical entries in dictionaries (e.g., Wikipedia) with prosaic definitions. We assumed that text encoders have internally captured the lexicon of Wikipedia pages during their training, and perform a better zero-shot semantic indexing with Wikipedia pages than on the gazetteer-based documents $D_{p, \textrm{geo}}$.
Hence, we extracted English descriptions $D_{p, \textrm{wiki}}$ from Wikipedia via their Wikipedia URL, denoted as the Wikipedia view (WikiView); we only extracted the sections \textit{Introduction}, \textit{Country}, and \textit{Geography}. We queried DBpedia to get the URL, and checked if the coordinates of the Wikipedia page are contained in the bounding box of the place. 

\boxpromptnumbered{WIKI}{\{Wikipedia content of p\}
}

For missing or rejected links, we first queried Nominatim to get the URL, or completed the template ultimately if the URL is missing:

\boxpromptnumbered{WIKI'}{\{place\} is located in \{parent\}.
\{place\} is surrounded by \{neighbor\}.
Coasts of \{place\} include \{waterbody\}.
}

\subsection{Language-model-based relevance function}
\label{section:methodology_lm_retrieval}

\partitle{Encoding mentions and entities.}
Let $\texttt{Enc}(\cdot)$ denote a text encoder that maps a text string into an embedding representation. For each entity $p \in P$, we pre-computed and cached the representation of its description with $\textrm{view} \in [ \textrm{geo}, \textrm{wiki} ]$:
\begin{equation}
    \mathbf{e}_p = \texttt{Enc}(D_{p, \textrm{view}})
\end{equation}

Given a spatial expression $t$ appearing in text $T$, we constructed a query string $Q(t,T)$ that highlights the mention and includes surrounding context, then computed from Section~\ref{section:benchmark_dataset}:
\begin{equation}
    \mathbf{q} = \texttt{Enc}\left( Q([t,T]) \right)
\end{equation}

We treated encoders as frozen during retrieval. 

\partitle{Similarity scoring}
We defined a generic similarity function $\mathrm{sim}(\mathbf{q}, \mathbf{e}_p)$ to score candidates. For single-vector encoders, the function is typically cosine similarity; for late-interaction models (e.g., ColBERT \cite{santhanamColBERTv2EffectiveEfficient2022}), the scoring follows their standard interaction mechanism.

\begin{figure}[hbt]
    \centering
    \includegraphics[width=1\columnwidth]{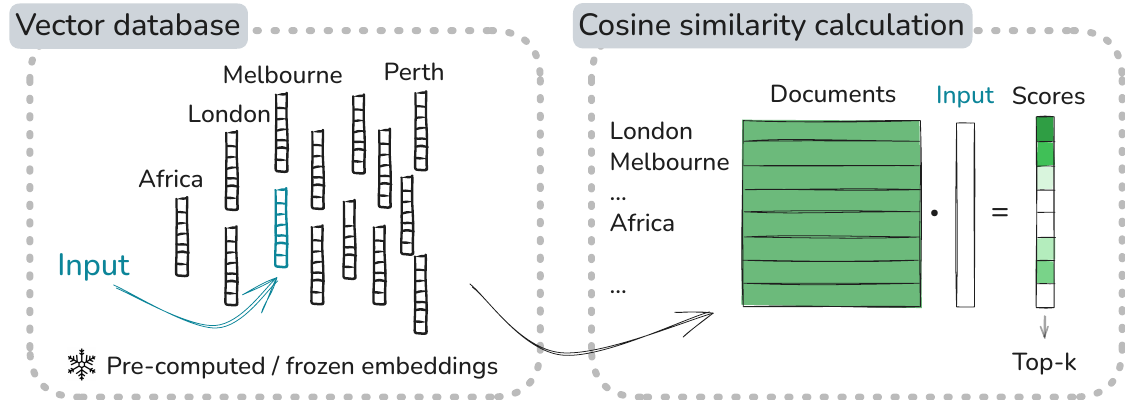}
    \caption{Brute-force search using all encoded documents $\textrm{Enc}(D_p)$ (\textcolor{ForestGreen}{green}).}
    \Description{Baseline \textcolor{red}{To remove}}
    \label{fig:baseline_model}
\end{figure}

\partitle{Unconstrained dense retrieval (Figure~\ref{fig:baseline_model}, Appendix~\ref{appendix:algo_brute_force})}
The brute-force approach ranks all candidates by embedding similarity and returns the top-1 entity:
\begin{equation}
\hat{p} = \arg\max_{p \in P} \; \mathrm{sim}(\mathbf{q}, \mathbf{e}_p).
\end{equation}
This setting tests whether encoder representations alone provide sufficient spatial fidelity without any explicit structural constraints.

\begin{figure*}[hbt]
    \centering
    \includegraphics[width=1.8\columnwidth]{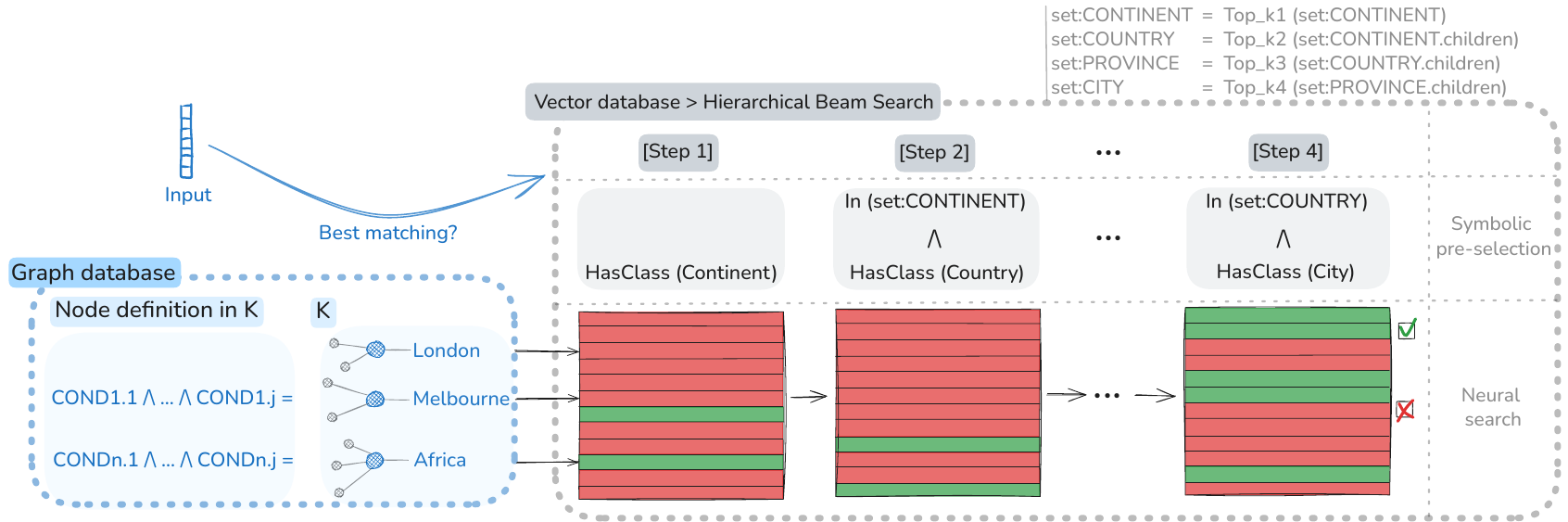}
    \caption{Hierarchical beam search (HBS). At Step $i$, we restrict candidates to those compatible with the current level and with the current beam's descendants: only a sub-set of candidates $\textrm{Enc}(D_p)$ is activated (\textcolor{ForestGreen}{green}) via a structured query.}
    \label{fig:hierarchical_model}
\end{figure*}

\partitle{Hierarchy-constrained retrieval (Figure~\ref{fig:hierarchical_model}, Appendix~\ref{appendix:algo_hbs}).}
To incorporate explicit spatial structure, we introduced a neuro-symbolic retrieval procedure that constrains and refines the candidate set using the gazetteer hierarchy encoded in $\mathcal{K}$. The core idea is to perform hierarchical beam search (HBS) from coarse geographic levels to finer ones: at each level, we keep a beam of the most similar candidates and expand only along valid hierarchical links (e.g., \texttt{subLevelOf}) to the next level.

Specifically, let $\ell_1 \rightarrow \ell_2 \rightarrow \cdots \rightarrow \ell_m$ denote an ordered sequence of levels (from coarse to fine). At step $i$, we restricted candidates to those compatible with level $\ell_i$ and with the current beam’s descendants in $\mathcal{K}$. We then ranked only this restricted set by similarity, kept the top-$k_i$ entities, and expanded to their children for the next step. The final prediction is the highest-scoring entity across all retained candidates $\bigcup_{i=1}^m \text{top-}k_i$, enabling backoff to coarser entities when finer predictions lack confidence.
This configuration is ablated with early stopping in Appendix~\ref{appendix:ablation}.

This procedure reduces confusion between geographically incompatible entities while preserving the encoder’s ability to interpret context and lexical variation: we \textbf{chain the embeddings by constraining the embedding space with symbolic filtering}.

\subsection{Evaluation metrics}
\label{section:eval}

Let $p^*$ be the ground-truth point location; $\hat{p}$ a
retrieved entity with footprint $\texttt{geom}(\hat{p})$ in $\mathcal{K}$ for the mention $t_i$; and the centroid $C(\hat{p}) = \texttt{centroid}(\texttt{geom}(\hat{p}))$. $\mathcal{K}$ has a maximum resolution at the city level.

\partitle{Point-based metrics (geocoding)} 
We report \textit{Acc@161km},
\textit{MED}, and \textit{AUC} based on the great-circle distance (denoted $\Delta_{\mathcal{K}}$).

\begin{definition}[Accuracy@161~km \cite{chengYouAreWhere2010}]
    \label{def:metric_acc_161}
    The Accuracy @ 161 km ($\text{Acc@}161$km) is defined as:
    \begin{equation}
        \text{Acc@}161~\text{km} = \frac{1}{N} \sum_{i=1}^N \mathbb{I}\left( \Delta_{\mathcal{K}}(C(\hat{p}_i), C(p_i^*)) \le 161~\text{km} \right)
    \end{equation}
    where $\mathbb{I}(\cdot)$ is the indicator function that returns $1$ if the condition $\Delta_{\mathcal{K}}(\hat{p}_i, p_i^*) \le 161~km$ is satisfied, and $0$ otherwise.
\end{definition}

\begin{definition}[Mean Error Distance (MED) \cite{grittaPragmaticGuideGeoparsing2020}]
    \label{def:metric_med}
    The Mean Error Distance ($\text{MED}$) is the arithmetic mean of the distances between each predicted entity $\hat{p}_i$ and its corresponding ground truth $p_i^*$:
    \begin{equation}
        \text{MED} = \frac{1}{N} \sum_{i=1}^N \Delta_{\mathcal{K}}( C ( \hat{p}_i ) , C ( p_i^*) )
    \end{equation}
\end{definition}



\begin{definition}[Area Under the Curve (AUC) \cite{jurgensGeolocationPredictionTwitter2015}]
\label{def:metric_auc}
Let the function $\varepsilon(i): [1, ..., N] \rightarrow [\varepsilon_1, \dots, \varepsilon_N]$ return the vector of ascending geocoding errors $\varepsilon_i = \max(\Delta_{\mathcal{K}}(\hat{p}_i, p_i^*), 1\text{~km})$. We define the AUC\footnote{A lower AUC is better. The logarithm degrades the score more for an error of 50~km for short distances rather than continental distances. The minimal error is 1~km to fix the $\ln(0)$ edge-case in the original paper \cite{jurgensGeolocationPredictionTwitter2015}.} \cite{jurgensGeolocationPredictionTwitter2015, grittaPragmaticGuideGeoparsing2020} via the discrete sum of log-scaled errors (with $d_{\text{max}}=20{,}039$~km) as:
\begin{equation}
    \text{AUC} = \frac{1}{N \ln(d_{\text{max}})} \sum_{i=1}^{N} \ln(\varepsilon_i)
\end{equation}
\end{definition}



\partitle{Coverage-based metric} In Figure~\ref{fig:example_opera_house}, the association between \textit{Sydney} and \textit{Australia} is spatially correct, but rejected by absolute point-based metrics. To account for indexation, we propose reporting relative accuracy by coverage:
\begin{gather}
    \text{Acc}_\text{coverage} = \frac{1}{N} \sum_{i=1}^N \mathbb{I}\left( \texttt{IsIn}(\hat{p}_i, p_i^*,h) \right)
    \label{eq:is_in}
\end{gather}
where 
$\texttt{IsIn}(\hat{p}, p^*,h) = \texttt{centroid}(p^*) \in \texttt{geom}(\hat{p}, h) $;
and $h = \textit{Country}$. 

In the previous metrics, we reported the top-1 prediction for each $\hat{p}_i$. In addition, we inspect the top-$k$ ($k \in [1, 3, 5, 10]$) predictions: we report the hit ratio HR$@k$ regarding the condition in Equation~\ref{eq:is_in}.

\section{Experimental setup}



\partitle{Text encoders.} We evaluated five widely used open embedding models for dense retrieval: all-MiniLM-L6-v2 (22M) \cite{reimersSentenceBERTSentenceEmbeddings2019}, ColBERTv2 (110M) \cite{santhanamColBERTv2EffectiveEfficient2022}, BGE base (109M) for English \cite{chenBGEM3EmbeddingMultiLingual2024a}, EmbeddingGemma (300M) \cite{veraEmbeddingGemmaPowerfulLightweight2025}, and Qwen3 Embeddings (0.6B) \cite{zhangQwen3EmbeddingAdvancing2025}.

\partitle{Baselines.} We compared all text encoders with BM25~\cite{robertsonProbabilisticRelevanceFramework2009, luBM25SOrdersMagnitude2024}, a standard strong keyword-based (lexical) retrieval model. 
Computational costs represent a limiting constraint in information retrieval, notably for spatial information applications \cite{huToponymResolutionLeveraging2024, cohnEvaluatingAbilityLarge2025a}. This constraint justifies the use of smaller models\footnote{e.g, below 1M parameters.} at early stages of the retrieval: we also compare against the lightweight models CamCoder \cite{grittaWhichMelbourneAugmenting2018} and MLG \cite{kulkarniMultiLevelGazetteerFreeGeocoding2021}, that are both gazetteer-free during inference. However, CamCoder and MLG are unfair comparisons as both models are trained to target exact locations (scores are reported but grayed out).

\partitle{Implementation details.} For HBS, beam widths across terrestrial hierarchy levels are set to $[3, 20, 50, 100, 1]$\footnote{Toponyms are restricted to terrestrial locations; maritime gazetteer matches are pruned and used solely to populate GeoView.}, retaining the top-500 candidates at each step.

\partitle{Gazetteer ambiguity.} The pruned gazetteer exhibits 17.1\% ambiguity: 10 countries drive roughly half the cases, while the remainder involve names shared across different countries (Appendix~\ref{appendix:toponym_ambiguity}).

(for 360 queries). 
\label{exp_overlap}

%% file: 00_CONTENT/08_Results.tex
\section{Results}





\input{00_CONTENT/FIGS_TABLES_IMG_ALG/fig_config_all_geocoding_HR_1}
\input{00_CONTENT/FIGS_TABLES_IMG_ALG/fig_distrib_top_k_all}

\parnone{} Figure~\ref{fig:geocoding_hr_config=1_1} and Figure~\ref{fig:geocoding_hr_config=2_1} compare brute-force search against HBS across GeoView (linearized triplets from $\mathcal{K}$) and WikiView (Wikipedia extracts), respectively. Across both views, point-based accuracy is consistently lower than coverage-based accuracy, with gaps of $\sim$0.1--0.2 for brute-force and $\sim$0.3 for HBS. This gap is expected: centroid distance strictly penalizes imprecise localization, whereas coverage-based evaluation accommodates spatially extended entities through spatial tolerance and coarse discretization. The two metrics are therefore complementary, reflecting localization precision versus geographic containment or plausibility.


Table~\ref{tab:acc_coverage_161} reports coverage-based accuracy scores. Except for Qwen3 on LGL, HBS improves all top-1 coverage-based accuracies, most notably for BGE and Gemma. Intuitively, coverage-based accuracy mainly reflects correct country-level detection. In Figures~\ref{fig:geocoding_hr_config=1_1} and~\ref{fig:geocoding_hr_config=2_1}, brute-force search correctly identifies the country in under 20\% of cases for ColBERTv2 and up to 70\% for Gemma; HBS increases these rates to 50\% and 90\%, respectively. Notably, on GV, BGE under GeoView jumps from 0.26 to 0.84, and ColBERTv2 under WikiView rises from 0.04 to 0.47.
HBS enables robust text-to-country grounding, which remains challenging under unconstrained retrieval. Explicit hierarchical constraints thus prevent coarse geographic mismatches driven purely by embedding similarity.

Figure~\ref{fig:distrib_top_k_config=kg} and Figure~\ref{fig:distrib_top_k_config=wiki} illustrate the distribution of country detection in the top-$k$ set via HR$@k$. HBS hits relevant countries more often at the highest ranks: the gap between the top-1 and top-10 is narrower with HBS than with brute-force search, dropping to less than 0.15 with Gemma on WikiView instead of 0.30.


Modern encoders (BGE, Qwen3, Gemma) outperform MiniLMv2 and ColBERTv2 under both brute-force search and HBS. However, overall performance remains modest ($<$0.6 for brute-force retrieval), showing that embedding similarity alone struggles to resolve spatial expressions. These results highlight a dual dependency: stronger encoders improve contextual semantics, but explicit spatial structure remains essential for reliable spatial-semantic indexing.





\parnone{}In our setting, the gazetteer's cut-off resolution prevents text encoders from acting as simple pattern matchers, forcing text encoders to generalize to the semantically closest concept. However, this conceptual match is not guaranteed to fall within 161~km of the ground-truth (and arbitrary) annotation. Table~\ref{tab:acc_161} reports point-based accuracy within a 161~km tolerance. As expected, HBS shows no consistent performance pattern, except for BGE across both GeoView and WikiView. Hence, $\text{Acc}@\text{161\,km}$ functions strictly as a binary threshold for fine-grained localization, and fails to capture error distributions beyond 161~km.

Additionally, we report Mean Error Distance (MED) and Area Under the Curve (AUC) respectively in Table~\ref{tab:comp_med} and Table~\ref{tab:comp_auc}. These global metrics evaluate whether spatial errors occur locally (<1{,}000~km), continentally (<4{,}000~km), or globally (>4{,}000--7{,}000~km).
On average, text encoder errors occur at a multi-country scale; notably, Gemma achieves an MED of 1,100~km on the GV dataset using HBS. This MED range is comparable to spatially fine-tuned models (CamCoder and MLG) with the exception of the WT dataset under MLG (MED of 200 km).

We observe that HBS tends to degrade MED and AUC, notably on the datasets LGL and WT: HBS cannot improve the retrieval of fine locations. One explanation is that text encoders lack robust representations for intermediate administrative boundaries, whereas some toponym resolutions are straightforward with a direct match via brute-force search. Consequently, routing beam search through poorly encoded intermediate regions potentially induces information loss rather than spatial refinement at final stages.

\parnone{}Figure~\ref{fig:geocoding_hr_config=1_1}, Figure~\ref{fig:geocoding_hr_config=2_1}, Figure~\ref{fig:distrib_top_k_config=kg} and Figure~\ref{fig:distrib_top_k_config=wiki} show that BM25 remains a competitive baseline, particularly when comparing to the earlier models MiniLMv2  and ColBERTv2. As BM25 works on keyword distributions, pattern matching remains a strong method in toponym resolution.
However, the more recent embedding encoders (e.g., BGE, Qwen3, and especially Gemma) achieve higher retrieval effectiveness, notably in terms of Acc@161~km; when combined with HBS, these recent models frequently match or exceed BM25. Indeed, natural language text possesses a richer contextual vocabulary beyond isolated toponyms that recent text encoders exploit.


 We observe major improvements when mapping a toponym to Wikipedia text (WikiView) instead of linearized gazetteer triplets (GeoView). We hypothesize that text encoders function as parametric lexical indexes for Wikipedia: during pre-training, their internal weights have effectively encode direct mappings between specific lexical surface forms and geographic entities. Consequently, WikiView performance reflects parametric memorization rather than true spatial generalization.

\input{00_CONTENT/FIGS_TABLES_IMG_ALG/tab_res_ACC_converage}

\input{00_CONTENT/FIGS_TABLES_IMG_ALG/tab_res_ACC}
\input{00_CONTENT/FIGS_TABLES_IMG_ALG/tab_res_MED}

\input{00_CONTENT/FIGS_TABLES_IMG_ALG/tab_res_AUC}




\input{00_CONTENT/FIGS_TABLES_IMG_ALG/fig_map_all_acc_entity_linking}


\parnone{}To understand where the system fails, Figure~\ref{fig:geocoding_mean_err_dist_kg@1} and Figure~\ref{fig:geocoding_mean_err_dist_wiki@1} map performance for a representative high-performing configuration (Gemma+HBS) for coverage-based accuracy across WikiView and GeoView. 
Errors are not spatially uniform: we observe clear geographic clusters of lower accuracy across datasets, notably in parts of the USA (<50\%) and western China (<25\%) for the datasets LGL and WT, primarily due to incorrect country-level retrieval.
This pattern suggests that failures are geographically structured rather than random. 

A likely contributing factor is higher toponym ambiguity in regions with many duplicated names (e.g., repeated European place names in the USA as presented in Appendix~\ref{appendix:toponym_ambiguity}), though isolating causes would require targeted analysis (e.g., stratifying errors by name ambiguity, entity density, and hierarchy depth). 
This ambiguity partly explains the high Mean Error Distance (MED) scores in Table~\ref{tab:comp_med}, driven by massive cross-continental misclassifications.




%% file: 00_CONTENT/FIGS_TABLES_IMG_ALG/fig_config_all_geocoding_HR_1.tex


\newcommand{\heightImg}{5.5cm} 
\newcommand{\heightLegend}{.50cm} 
\newcommand{\spaceHz}{-2ex} 

\begin{figure*}[htb] 
    \centering 
    \subfigure[MiniLMv2]{ 
        \resizebox*{!}{\heightImg}{\includegraphics{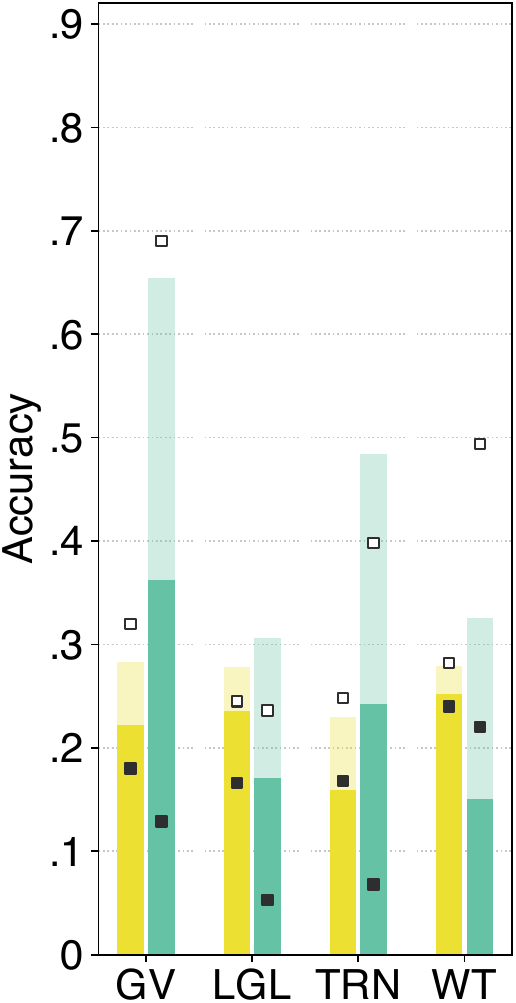}} 
        \label{fig:geocoding_acc_sbert} 
    } 
    \hspace{\spaceHz} 
    \subfigure[ColBERTv2]{ 
        \resizebox*{!}{\heightImg}{\includegraphics{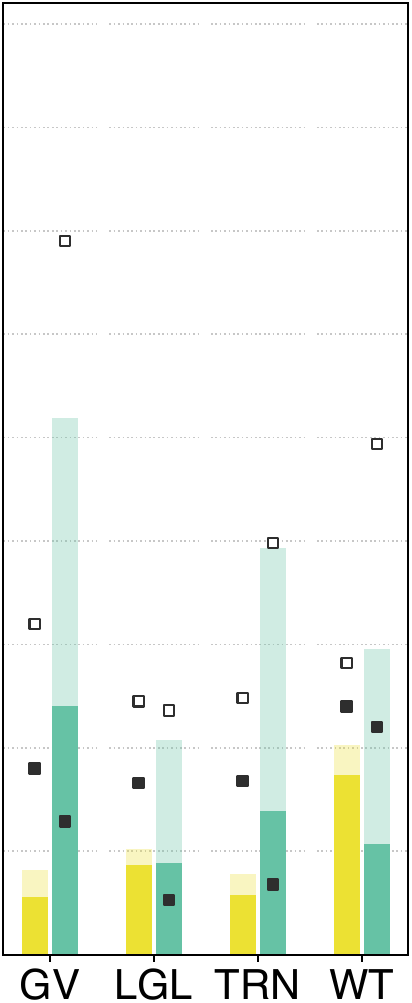}} 
        \label{fig:geocoding_acc_colbert} 
    } 
    \hspace{\spaceHz} 
    \subfigure[BGE base]{ 
        \resizebox*{!}{\heightImg}{\includegraphics{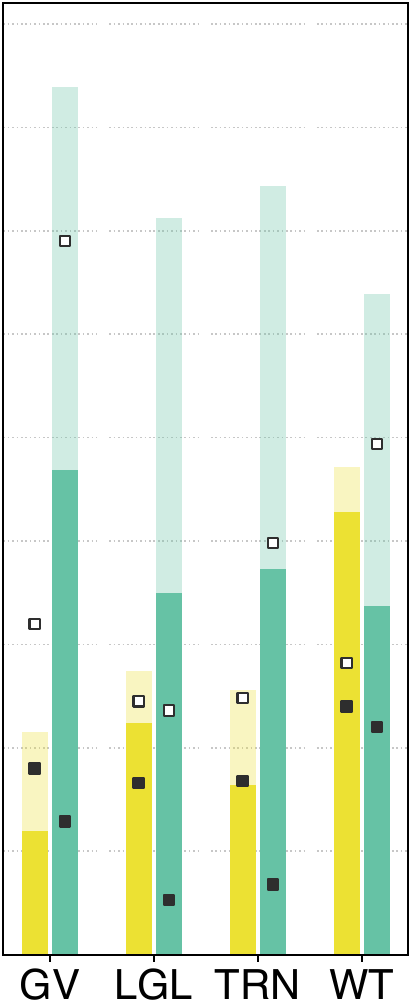}} 
        \label{fig:geocoding_acc_bge} 
    } 
    \hspace{\spaceHz} 
    \subfigure[Qwen3]{ 
        \resizebox*{!}{\heightImg}{\includegraphics{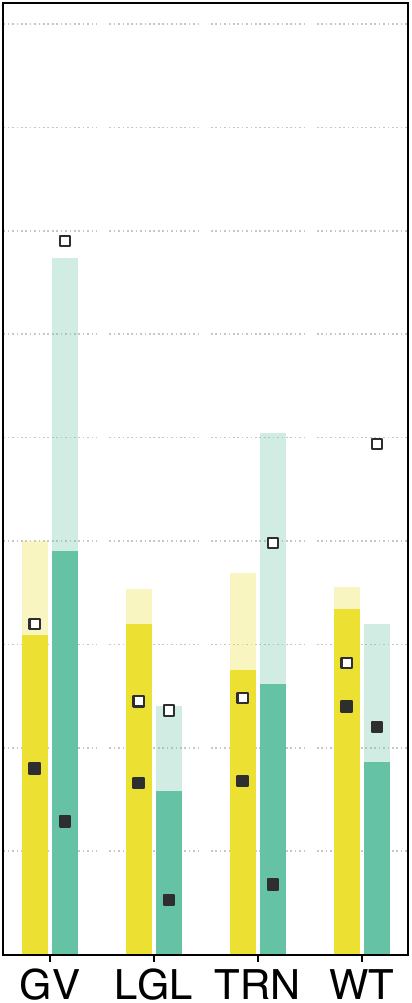}} 
        \label{fig:geocoding_acc_qwen} 
    } 
    \hspace{\spaceHz} 
    \subfigure[Gemma]{ 
        \resizebox*{!}{\heightImg}{\includegraphics{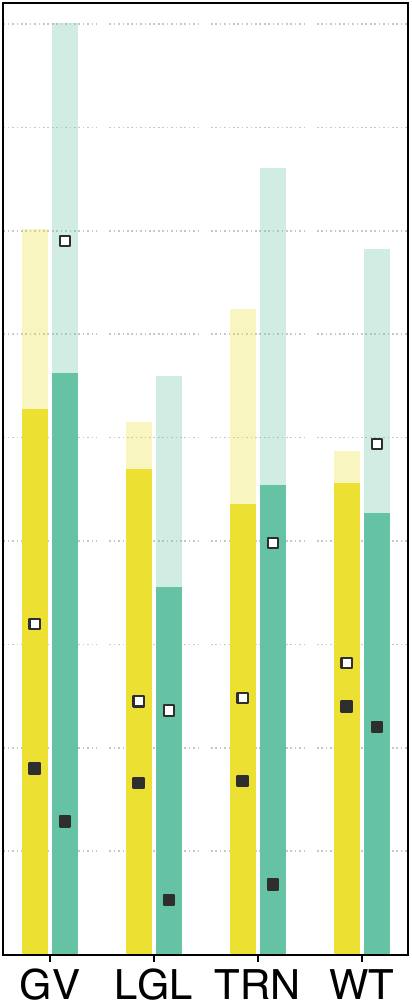}} 
        \label{fig:geocoding_acc_gemma} 
    } 
    \smallskip
    \\ 
    \subfigure{ 
        \resizebox*{!}{\heightLegend}{ 
            \includegraphics{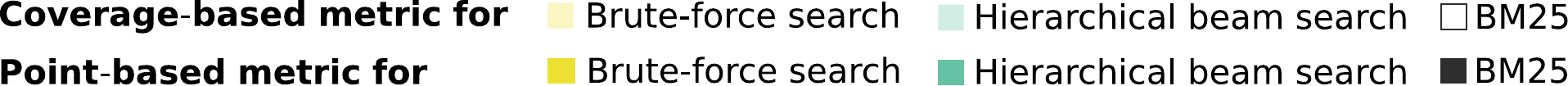} 
        } 
    } 

    \caption{Accuracy of spatial-semantic indexing on the configuration GeoView. Models are sorted by release date.} 
    \label{fig:geocoding_hr_config=1_1}

    \subfigure[MiniLMv2]{ 
        \resizebox*{!}{\heightImg}{\includegraphics{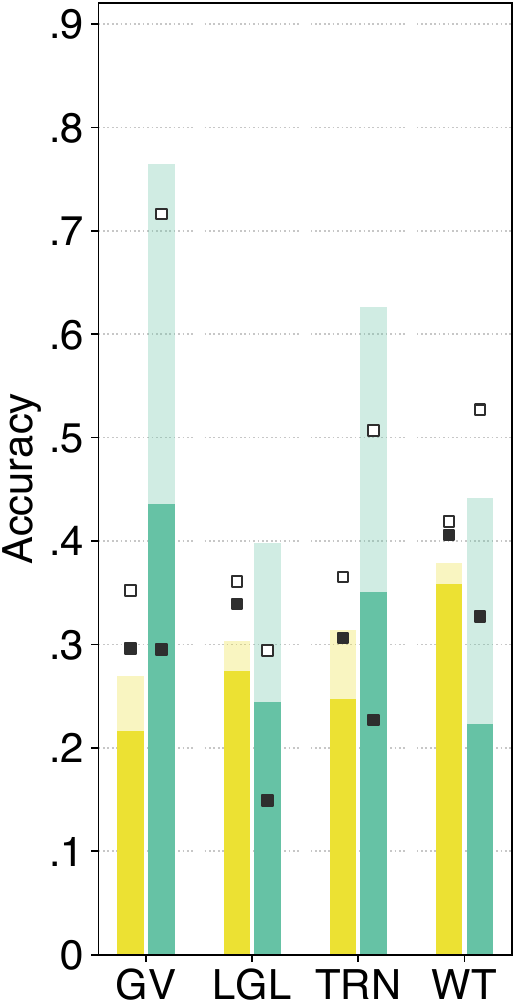}} 
        \label{fig:geocoding_acc_sbert} 
    } 
    \hspace{\spaceHz} 
    \subfigure[ColBERTv2]{ 
        \resizebox*{!}{\heightImg}{\includegraphics{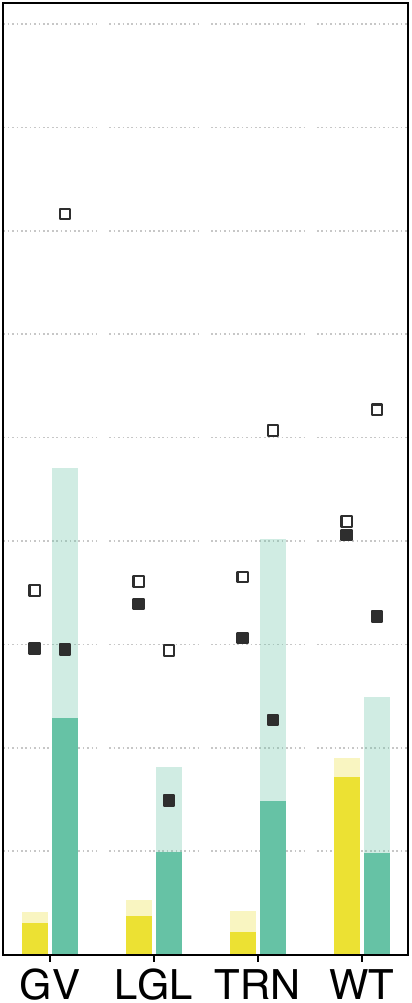}} 
        \label{fig:geocoding_acc_colbert} 
    } 
    \hspace{\spaceHz} 
    \subfigure[BGE base]{ 
        \resizebox*{!}{\heightImg}{\includegraphics{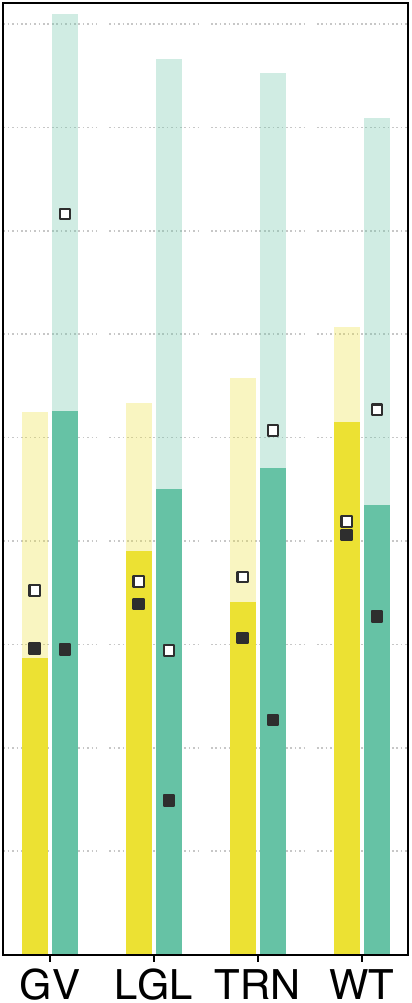}} 
        \label{fig:geocoding_acc_bge} 
    } 
    \hspace{\spaceHz} 
    \subfigure[Qwen3]{ 
        \resizebox*{!}{\heightImg}{\includegraphics{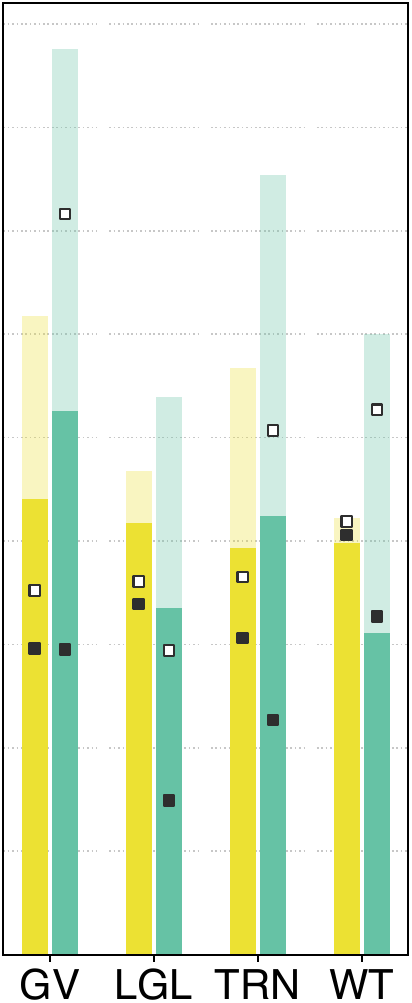}} 
        \label{fig:geocoding_acc_qwen} 
    } 
    \hspace{\spaceHz} 
    \subfigure[Gemma]{ 
        \resizebox*{!}{\heightImg}{\includegraphics{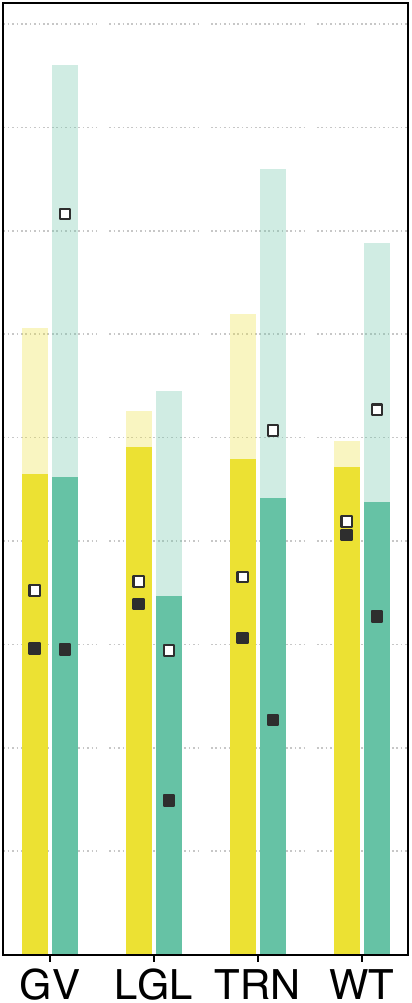}} 
        \label{fig:geocoding_acc_gemma} 
    } 
    \smallskip
    \\ 
    \subfigure{ 
        \resizebox*{!}{\heightLegend}{ 
            \includegraphics{01_IMAGES/stats_eval_metric=HR_161km_1_topk=top_k_legend.pdf} 
        } 
    } 

    \caption{Accuracy of spatial-semantic indexing on the configuration WikiView. Models are sorted by release date.} 
    \label{fig:geocoding_hr_config=2_1} 
\end{figure*}

%% file: 00_CONTENT/FIGS_TABLES_IMG_ALG/fig_distrib_top_k_all.tex
\newcommand{\heightImgStart}{3.25cm} 
\renewcommand{\heightImg}{3.2cm} 
\renewcommand{\heightLegend}{.37cm} 
\renewcommand{\spaceHz}{-2ex}
\newcommand{\colorsquare}[1]{%
    \setlength{\fboxrule}{0.5pt}
    \setlength{\fboxsep}{0pt}
    \fcolorbox{black}{#1}{\rule{0pt}{4pt}\rule{4pt}{0pt}}%
}

\newcommand{\colorcircle}[1]{%
    \setlength{\unitlength}{1pt}%
    \begin{picture}(6,6)
        \put(3,2){\color{#1}\circle*{4}} 
        \put(3,2){\circle{4}}            
    \end{picture}%
}

\renewcommand{\heightLegend}{.37cm} 
\renewcommand{\spaceHz}{-2ex} 
\begin{figure}[htb] 
    \centering 
    \subfigure[BM25]{ 
        \resizebox*{!}{\heightImgStart}{\includegraphics{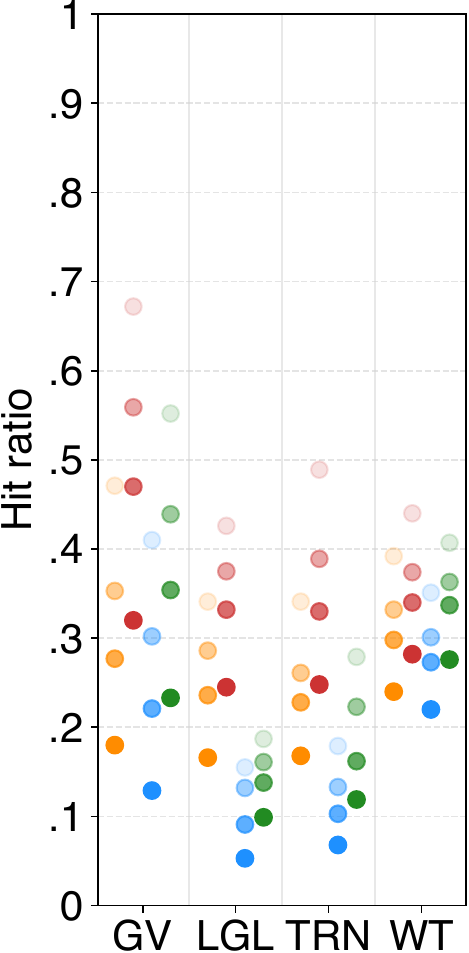}} 
        \label{fig:geocoding_acc_sbert} 
    } 
    \hspace{\spaceHz} 
    \subfigure[MiniLM]{ 
        \resizebox*{!}{\heightImg}{\includegraphics{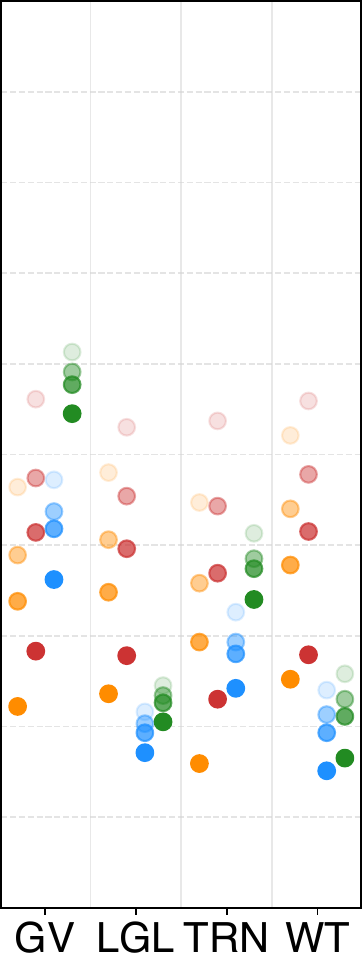}} 
        \label{fig:geocoding_acc_colbert} 
    } 
    \hspace{\spaceHz} 
    \subfigure[ColBERT]{ 
        \resizebox*{!}{\heightImg}{\includegraphics{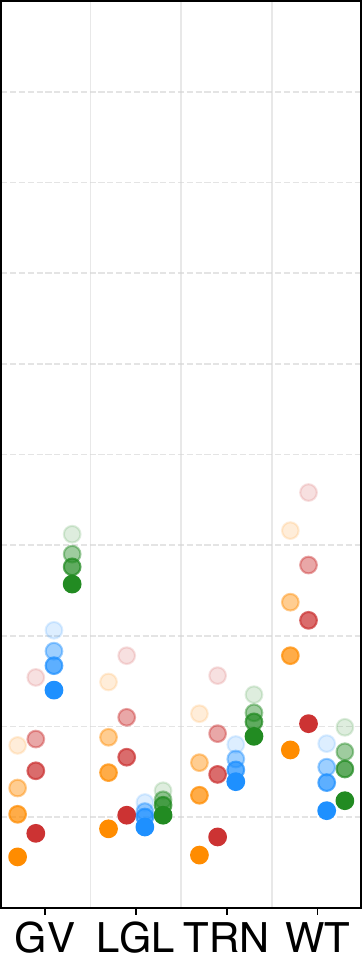}} 
        \label{fig:geocoding_acc_bge} 
    } 
    \hspace{\spaceHz} 
    \subfigure[BGE]{ 
        \resizebox*{!}{\heightImg}{\includegraphics{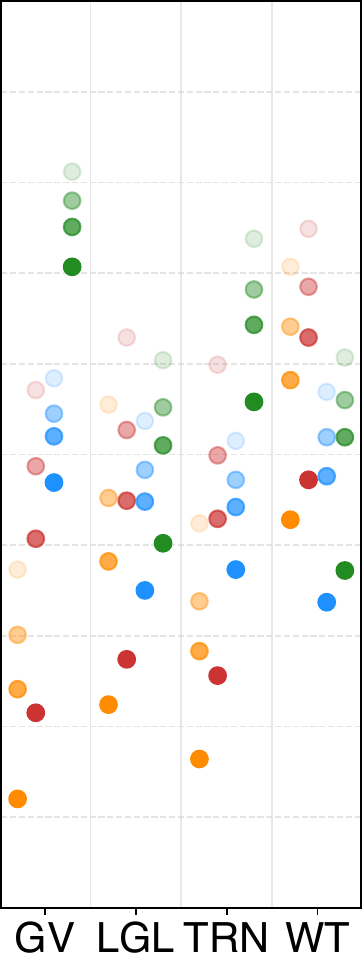}} 
        \label{fig:geocoding_acc_qwen} 
    } 
    \hspace{\spaceHz} 
    \subfigure[Qwen3]{ 
        \resizebox*{!}{\heightImg}{\includegraphics{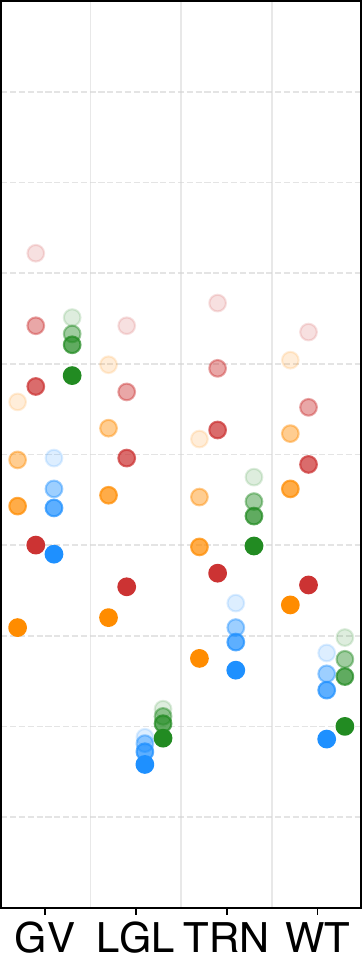}} 
        \label{fig:geocoding_acc_gemma} 
    } 
    \hspace{\spaceHz} 
    \subfigure[Gemma]{ 
        \resizebox*{!}{\heightImg}{\includegraphics{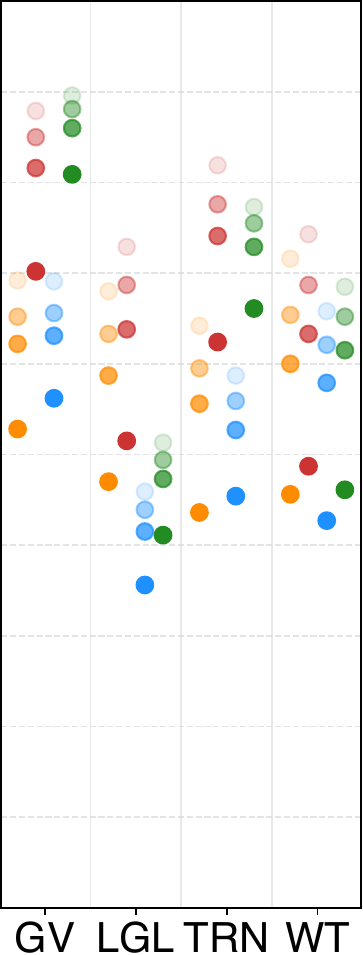}} 
        \label{fig:geocoding_acc_gemma} 
    }
    \caption{HR$_k$ ($k \in [1,3,5,10]$ in the configuration GeoView for the brute-force search (point \colorcircle{Dandelion} / coverage \colorcircle{Mahogany}) and HBS (point \colorcircle{Cyan} / coverage \colorcircle{ForestGreen}). Higher $k$ leads to higher HR.} 
    \label{fig:distrib_top_k_config=kg} 

    \subfigure[BM25]{ 
        \resizebox*{!}{\heightImgStart}{\includegraphics{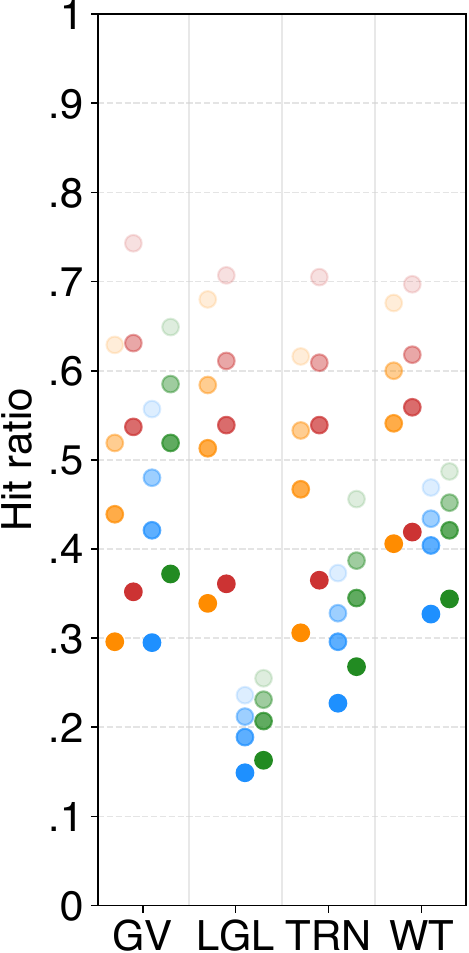}} 
        \label{fig:geocoding_acc_sbert} 
    } 
    \hspace{\spaceHz} 
    \subfigure[MiniLM]{ 
        \resizebox*{!}{\heightImg}{\includegraphics{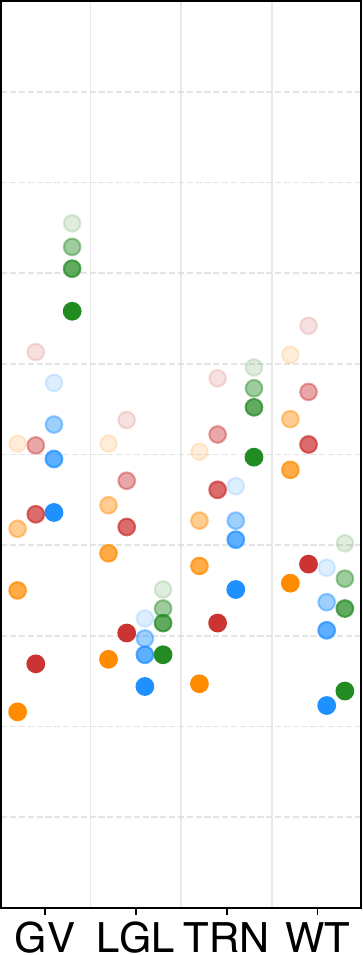}} 
        \label{fig:geocoding_acc_colbert} 
    } 
    \hspace{\spaceHz} 
    \subfigure[ColBERT]{ 
        \resizebox*{!}{\heightImg}{\includegraphics{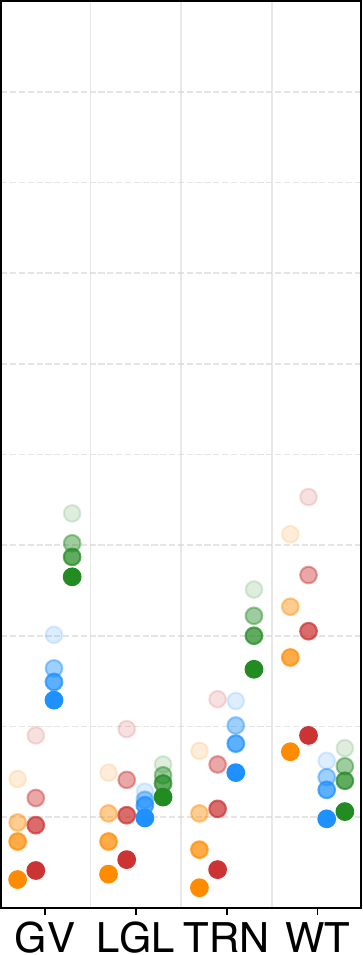}} 
        \label{fig:geocoding_acc_bge} 
    } 
    \hspace{\spaceHz} 
    \subfigure[BGE]{ 
        \resizebox*{!}{\heightImg}{\includegraphics{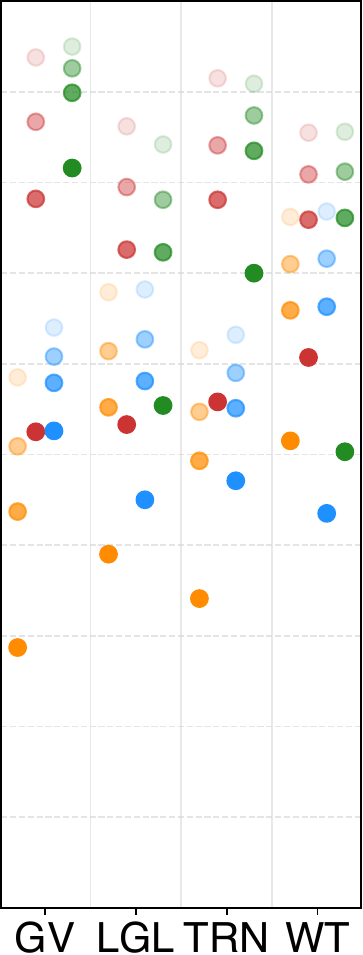}} 
        \label{fig:geocoding_acc_qwen} 
    } 
    \hspace{\spaceHz} 
    \subfigure[Qwen3]{ 
        \resizebox*{!}{\heightImg}{\includegraphics{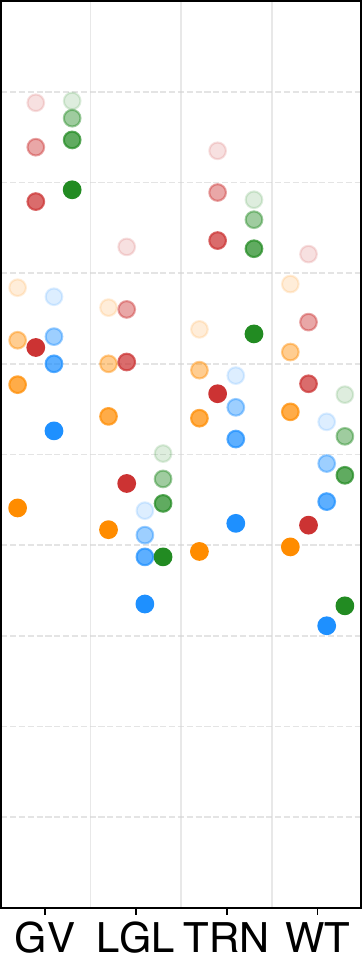}} 
        \label{fig:geocoding_acc_gemma} 
    } 
    \hspace{\spaceHz} 
    \subfigure[Gemma]{ 
        \resizebox*{!}{\heightImg}{\includegraphics{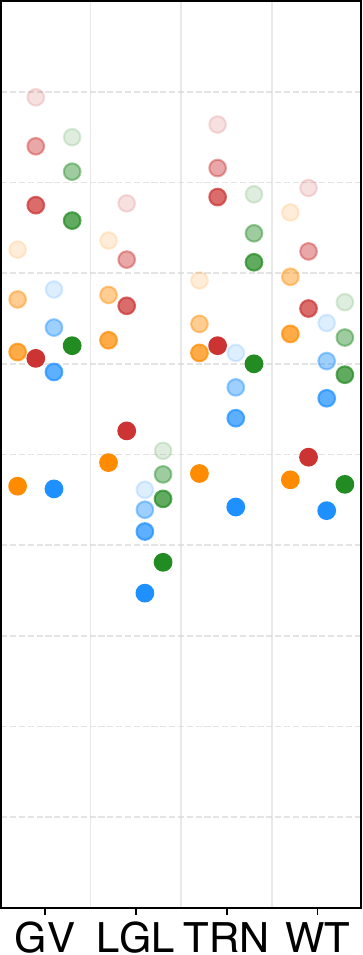}} 
        \label{fig:geocoding_acc_gemma} 
    }
    \caption{HR$_k$ ($k \in [1,3,5,10]$ in the configuration WikiView for the brute-force search (point \colorcircle{Dandelion} / coverage \colorcircle{Mahogany}) and HBS (point \colorcircle{Cyan} / coverage \colorcircle{ForestGreen}). Higher $k$ leads to higher HR.} 
    \label{fig:distrib_top_k_config=wiki} 
\end{figure}

%% file: 00_CONTENT/FIGS_TABLES_IMG_ALG/tab_res_ACC_converage.tex
\begin{table}[hbt]
  \caption{Comparison for Acc$_{\text{coverage}}$ ($\uparrow$). Best for the configuration noted as \textbf{.90}; and best for the dataset as \best{.91} for example.}
  \label{tab:comp_0}
  \centering
  {\Small
  \begin{tabular}{llcccc|cccc}
    \toprule
    & & \multicolumn{4}{c}{Brute-force search} & \multicolumn{4}{c}{Hierarchical beam search} \\
    View & Model & GV & LGL & TRN & WT & GV & LGL & TRN & WT\\
    \midrule
    \multirow{6}{*}{Geo} & BM25 & .32 & .25 & .25 & .28  & .69 & .24 & .40 & .49 \\
    & MiniLMv2 & .28 & .28 & .23 & .28 & .65 & .31 & .48 & .33 \\
    & ColBERTv2 & .08 & .10 & .08 & .20 & .52 & .21 & .39 & .30 \\
    & BGE base & .26 & .27 & .26 & .47 & .84 & \textbf{.71} & .74 & .64 \\
    & Qwen3 & .40 & .35 & .37 & .36  & .67 & .24 & .50 & .32 \\
    & Gemma & .70 & .52 & .62 & .49 & \textbf{.90} & .56 & \textbf{.76} & \textbf{.68} \\
    \midrule
    \multirow{6}{*}{Wiki} & BM25 & .35 & .36 & .37 & .42 & .72 & .29 & .51 & .53 \\
    & MiniLMv2 & .27 & .30 & .31 & .38 & .77 & .40 & .63 & .44 \\
    & ColBERTv2 & .04 & .05 & .04 & .19 & .47 & .18 & .40 & .25 \\
    & BGE base & .53 & .53 & .56 & .61 & \best{.91} & \best{.87} & \best{.85} & \best{.81} \\
    & Qwen3 & .62 & .47 & .57 & .42 & .88 & .54 & .75 & .60 \\
    & Gemma & .61 & .53 & .62 & .50 & .86 & .55 & .76 & .69 \\
    \bottomrule
  \end{tabular}
  }
  \label{tab:acc_coverage_161}
\end{table}

%% file: 00_CONTENT/FIGS_TABLES_IMG_ALG/tab_res_ACC.tex
\begin{table}[hbt]
  \caption{Comparison for Acc@161km ($\uparrow$).}
  \label{tab:comp_0}
  \centering
  {\Small
  \begin{NiceTabular}{llcccc|cccc}
    \CodeBefore
      \tikz \fill [pattern=north east lines, pattern color=gray!80] (3-|7) rectangle (5-|11);
    \Body
    \toprule
    & & \multicolumn{4}{c}{Brute-force search} & \multicolumn{4}{c}{Hierarchical beam search} \\
    View & Model & GV & LGL & TRN & WT & GV & LGL & TRN & WT\\
    \midrule
    & \opt{CamCoder}  & \opt{\textbf{.82}} & \opt{\textbf{.76}} & \opt{\textbf{.67}} & \opt{.65}  &  &  &  &  \\ 
    & \opt{MLG}  & \opt{.49} & \opt{.53} & \opt{-} & \opt{\textbf{.91}}  &  &  &  &  \\
    \midrule
    \multirow{6}{*}{Geo} & BM25 & .18 & .17 & .17 & .24 & .13 & .05 & .07 & .22 \\
    & MiniLMv2 & .22 & .24 & .16 & .25 & .36 & .17 & .24 & .15 \\
    & ColBERTv2 & .06 & .09 & .06 & .17 & .24 & .09 & .14 & .11 \\
    & BGE base & .12 & .22 & .16 & .43 & .47 & .35 & .37 & .34 \\
    & Qwen3 & .31 & .32 & .28 & .33 & .39 & .16 & .26 & .19 \\
    & Gemma & .53 & \textbf{.47} & .44 & \textbf{.46} & \best{.56} & .36 & \textbf{.45} & .43 \\
    \midrule
    \multirow{6}{*}{Wiki} & BM25 & .30 & .34 & .31 & .41 & .29 & .15 & .23 & .33 \\
    & MiniLMv2 & .22 & .27 & .25 & .36 & .44 & .24 & .35 & .22 \\
    & ColBERTv2 & .03 & .04 & .02 & .17 & .23 & .10 & .15 & .10 \\
    & BGE base & .29 & .39 & .34 & .52 & \textbf{.53} & .45 & .47 & .43 \\
    & Qwen3 & .44 & .42 & .39 & .40 & \textbf{.53} & .34 & .42 & .31 \\
    & Gemma & .48 & \best{.49} & \best{.48} & \best{.47} & .46 & .35 & .44 & .44 \\
    \bottomrule
  \end{NiceTabular}
  \label{tab:acc_161}
  }
\end{table}

%% file: 00_CONTENT/FIGS_TABLES_IMG_ALG/tab_res_MED.tex
\begin{table}[hbt]
  \caption{Comparison of MED ($\downarrow$). 7{,}000~km roughly represents the distance between the USA and Europe; 4{,}000~km between the west and east coasts of the USA; 1{,}000~km between the Western and Eastern coasts of Spain.}
  \label{tab:comp_med}
  \centering
  {\Small
  \begin{NiceTabular}{llcccc|cccc}
    \CodeBefore
      \tikz \fill [pattern=north east lines, pattern color=gray!80] (3-|7) rectangle (5-|11);
    \Body
    \toprule
    \multicolumn{2}{c}{($\times100$~km)} & \multicolumn{4}{c}{Brute-force search} & \multicolumn{4}{c}{Hierarchical beam search} \\
    View & Model & GV & LGL & TRN & WT & GV & LGL & TRN & WT \\
    \midrule
    & \opt{CamCoder}  & \opt{\textbf{7}} & \opt{22} & \opt{\textbf{30}} & \opt{11}  &  &  &  &  \\ 
    & \opt{MLG} & \opt{17} & \opt{\textbf{14}} & \opt{-} & \opt{\textbf{2}}  &  &  &  &  \\
    \midrule
    \multirow{6}{*}{Geo} & BM25 & 24 & 43 & 37 & 47 & 31 & 59 & 53 & 48 \\
    & MiniLMv2 & 26 & 21 & 26 & 33 & 28 & 61 & 42 & 57 \\
    & ColBERTv2 & 46 & 41 & 44 & 33 & 37 & 63 & 47 & 52 \\
    & BGE base & 36 & \textbf{18} & 30 & \textbf{20} & 14 & 25 & 22 & 30 \\
    & Qwen3 & 23 & 23 & 26 & 30 & 25 & 66 & 41 & 52 \\
    & Gemma & 12 & \textbf{18} & \textbf{17} & 23 & \textbf{11} & 38 & 19 & 27 \\
    \midrule
    \multirow{6}{*}{Wiki} & BM25 & 34 & 34 & 30 & 33 & 27 & 66 & 46 & 44 \\
    & MiniLMv2 & 35 & 35 & 35 & 35 & 20 & 51 & 32 & 47 \\
    & ColBERTv2 & 56 & 50 & 51 & 38 & 39 & 59 & 48 & 53 \\
    & BGE base & 18 & \best{16} & 22 & \best{17} & \best{10} & \best{16} & \best{16} & 20 \\
    & Qwen3 & 16 & 23 & 22 & 28 & 11 & 38 & 22 & 34 \\
    & Gemma & 15 & 21 & 17 & 24 & 13 & 39 & 21 & 27 \\
    \bottomrule
  \end{NiceTabular}
  }
\end{table}

%% file: 00_CONTENT/FIGS_TABLES_IMG_ALG/tab_res_AUC.tex
\begin{table}[hbt]
  \caption{Comparison for AUC ($\downarrow$).}
  \label{tab:comp_auc}
  \centering
  {\Small
  \begin{NiceTabular}{llcccc|cccc}
    \CodeBefore
      \tikz \fill [pattern=north east lines, pattern color=gray!80] (3-|7) rectangle (5-|11);
    \Body
    \toprule
    & & \multicolumn{4}{c}{Brute-force search} & \multicolumn{4}{c}{Hierarchical beam search} \\
    View & Model & GV & LGL & TRN & WT & GV & LGL & TRN & WT \\
    \midrule
    & \opt{CamCoder}  & \opt{.65} & \opt{.60} & \opt{\textbf{.31}} & \opt{.49}  &  &  &  &  \\ 
    & \opt{MLG}  & \opt{\textbf{.55}} & \opt{\textbf{.54}} & \opt{-} & \opt{\textbf{.37}}  &  &  &  &  \\
    \midrule
    \multirow{6}{*}{Geo} & BM25 & .68 & .73 & .72 & .72 & .71 & .82 & .79 & .73 \\
    & MiniLMv2 & .67 & .65 & .68 & .67 & .63 & .78 & .71 & .77 \\
    & ColBERTv2 & .78 & .76 & .77 & .70 & .70 & .82 & .76 & .78 \\
    & BGE base & .74 & .64 & .70 & \textbf{.58} & .56 & .62 & .60 & .64 \\
    & Qwen3 & .63 & .62 & .65 & .63 & .61 & .80 & .69 & .75 \\
    & Gemma & .53 & \best{.56} & \textbf{.57} & \textbf{.58} & \best{.52} & .66 & \textbf{.57} & .60 \\
    \midrule
    \multirow{6}{*}{Wiki} & BM25 & .66 & .65 & .65 & .62 & .64 & .80 & .72 & .68 \\
    & MiniLMv2 & .69 & .67 & .69 & .63 & .58 & .73 & .65 & .72 \\
    & ColBERTv2 & .82 & .80 & .81 & .71 & .70 & .81 & .75 & .79 \\
    & BGE base & .63 & .59 & .62 & \best{.54} & .53 & .57 & \best{.56} & .57 \\
    & Qwen3 & .57 & .59 & .60 & .61 & \textbf{.53} & .67 & .59 & .65 \\
    & Gemma & .55 & \best{.56} & \best{.56} & .57 & \textbf{.55} & .67 & .58 & .59 \\
    \bottomrule
  \end{NiceTabular}
  }
\end{table}

%% file: 00_CONTENT/FIGS_TABLES_IMG_ALG/fig_map_all_acc_entity_linking.tex


\newcommand{\imgWidthLeft}{0.93}
\newcommand{\imgWidthRight}{0.98}
\newcommand{\hshort}{-0ex}
\newcommand{\vshort}{-2exex}

\begin{figure*}[htb]
    \centering
    \subfigure[GeoVirus]{
    \resizebox*{\imgWidthLeft\columnwidth}{!}{
        \includegraphics{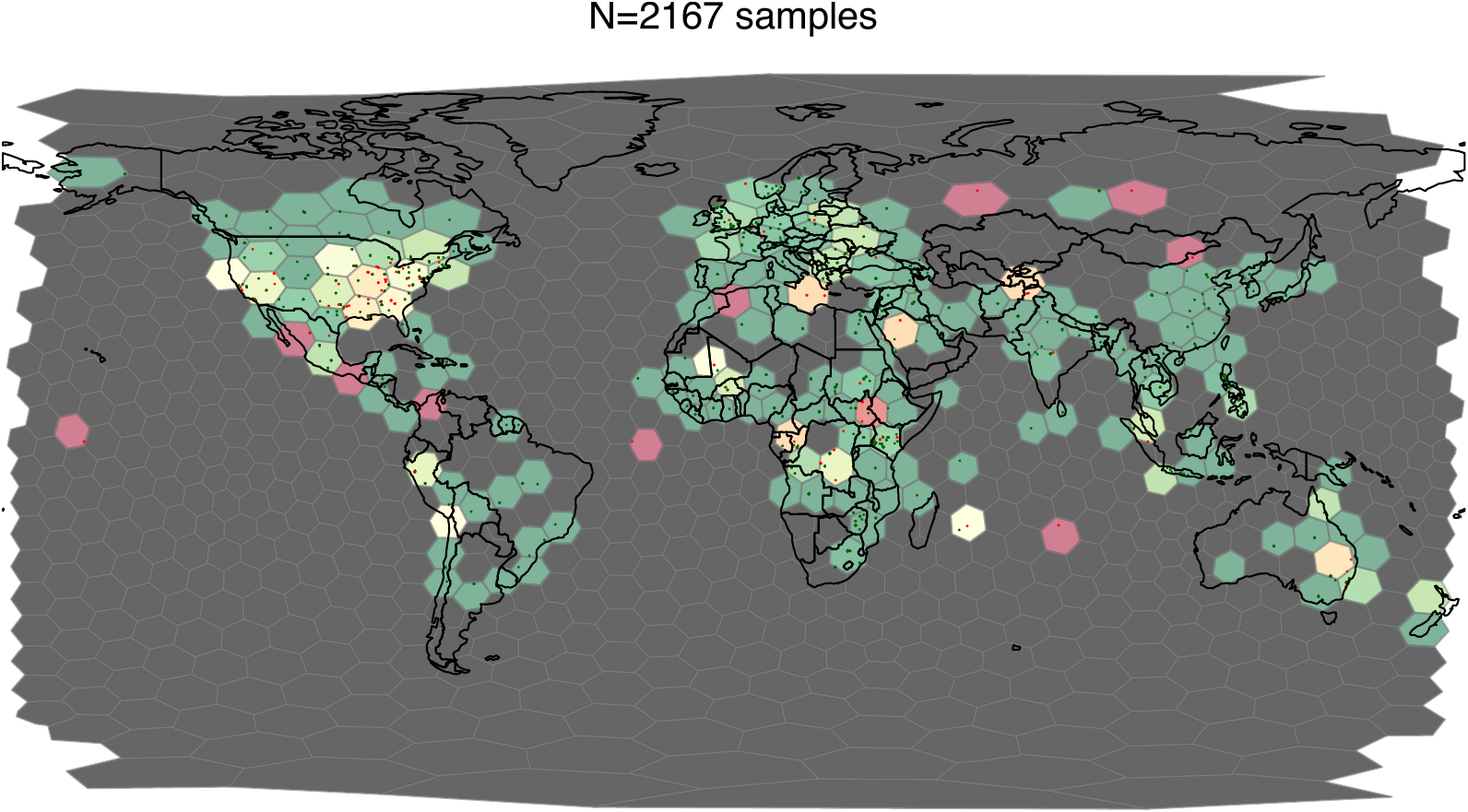} 
    }
    \label{fig:map_kg_acc_ent_link_geovirus}
     }
    \hspace{\hshort}
    \subfigure[LGL]{
    \resizebox*{\imgWidthRight\columnwidth}{!}{
        \includegraphics{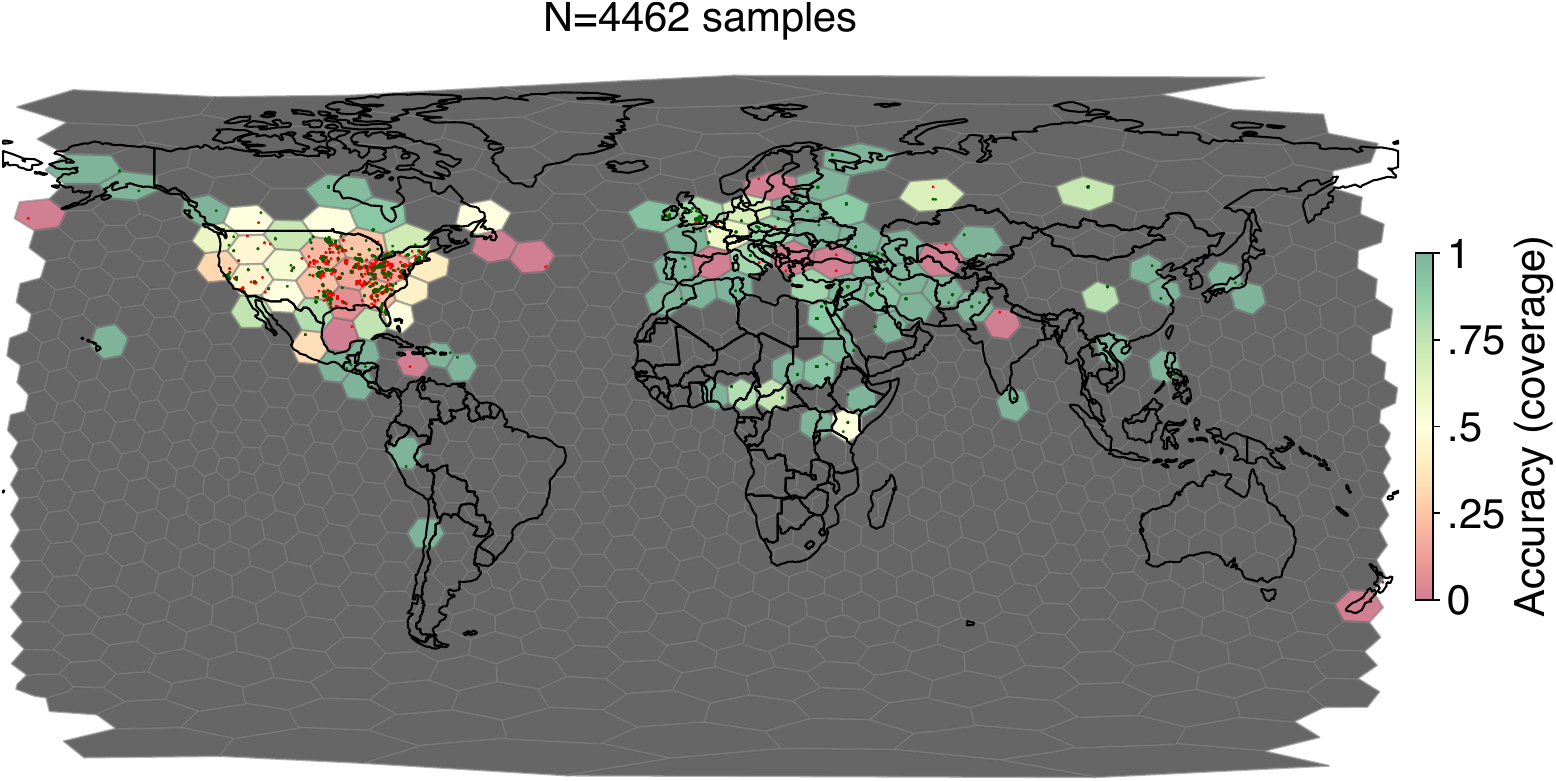}
    }
    \label{fig:map_kg_acc_ent_link_lgl}
    }
    \hspace{\hshort}
    \subfigure[TR-News]{
    \resizebox*{\imgWidthLeft\columnwidth}{!}{
        \includegraphics{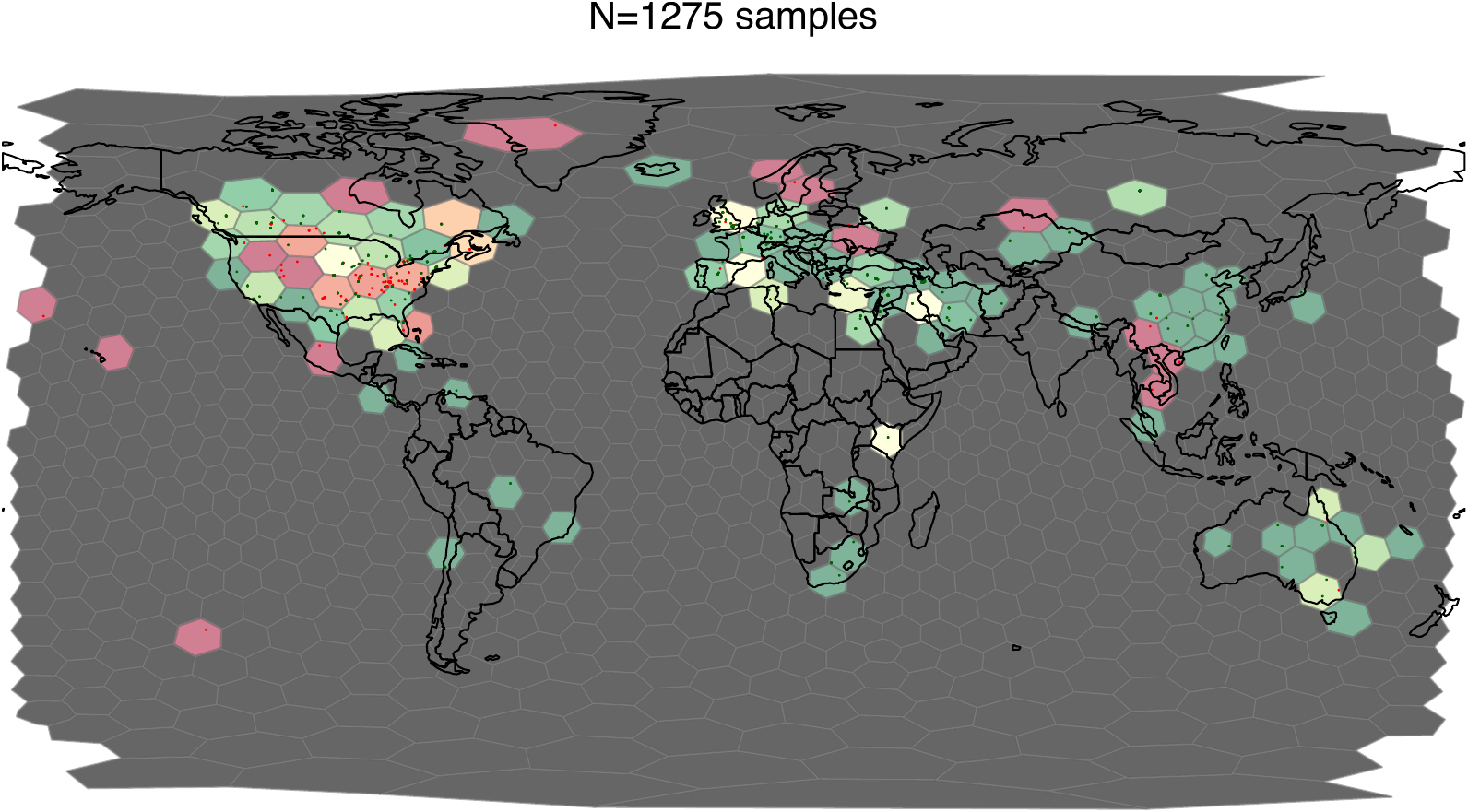}
    }
    \label{fig:map_kg_acc_ent_link_trnews}
    }   
    \hspace{\hshort}
    \subfigure[WikTor]{
    \resizebox*{\imgWidthRight\columnwidth}{!}{
        \includegraphics{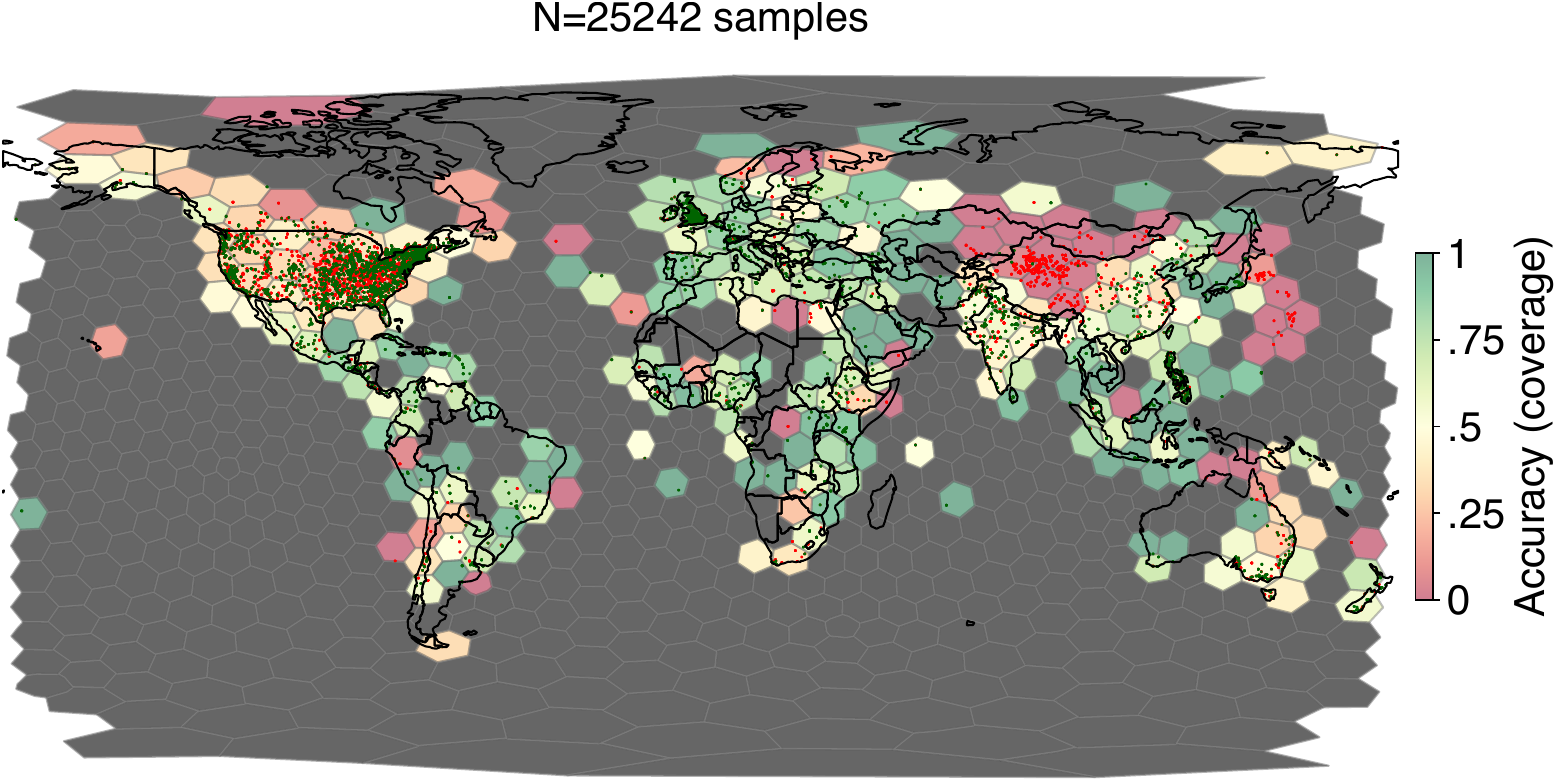}
    }
    \label{fig:map_kg_acc_ent_link_wiktor}
    }   
    \caption{Local Acc$_{\textrm{coverage}}$ (Gemma with HBS) in the GeoView.} 
    \label{fig:geocoding_mean_err_dist_kg@1}

    \subfigure[GeoVirus]{
    \resizebox*{\imgWidthLeft\columnwidth}{!}{
        \includegraphics{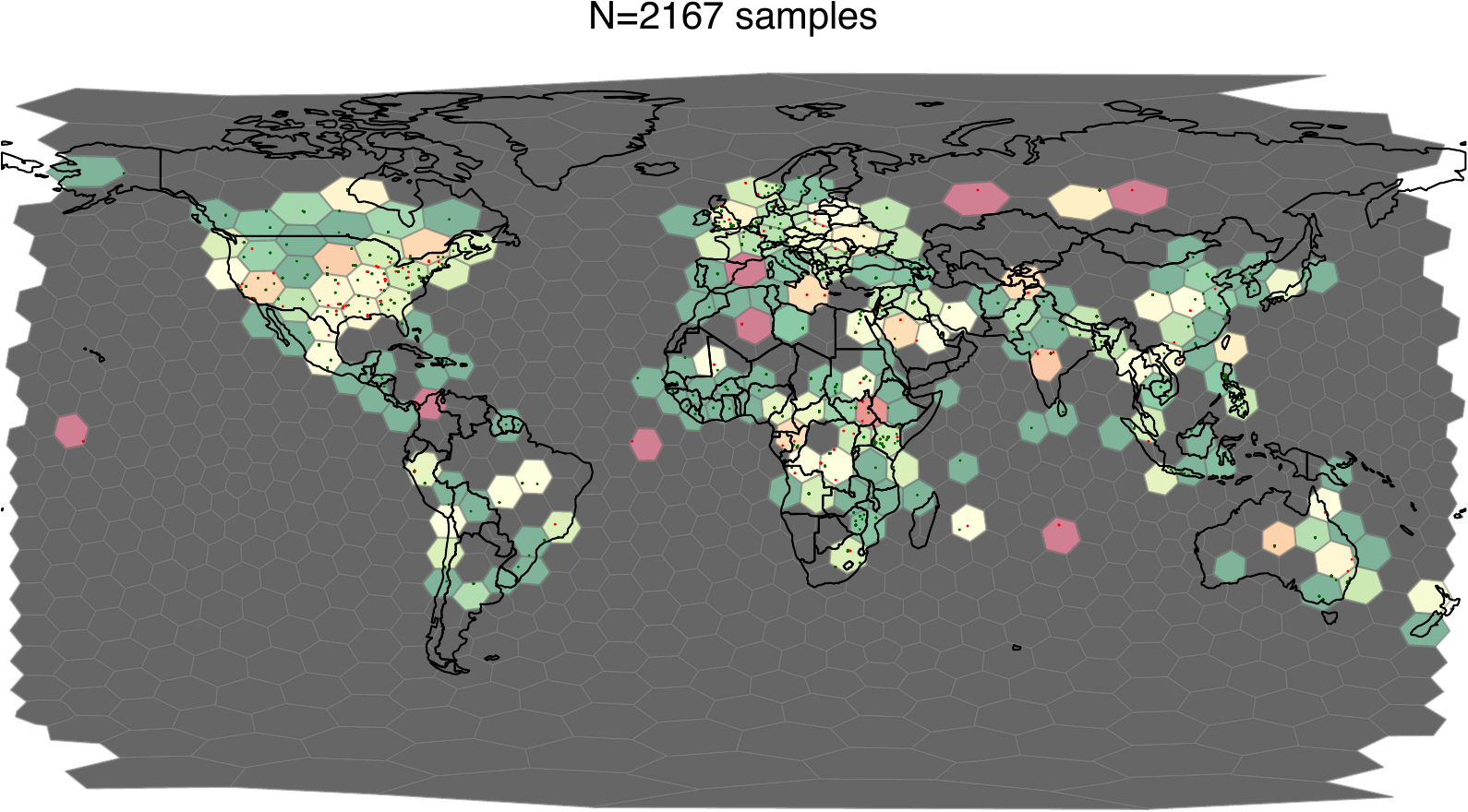} 
    }
    \label{fig:map_wiki_acc_ent_link_geovirus}
     }
    \hspace{\hshort}
    \subfigure[LGL]{
    \resizebox*{\imgWidthRight\columnwidth}{!}{
        \includegraphics{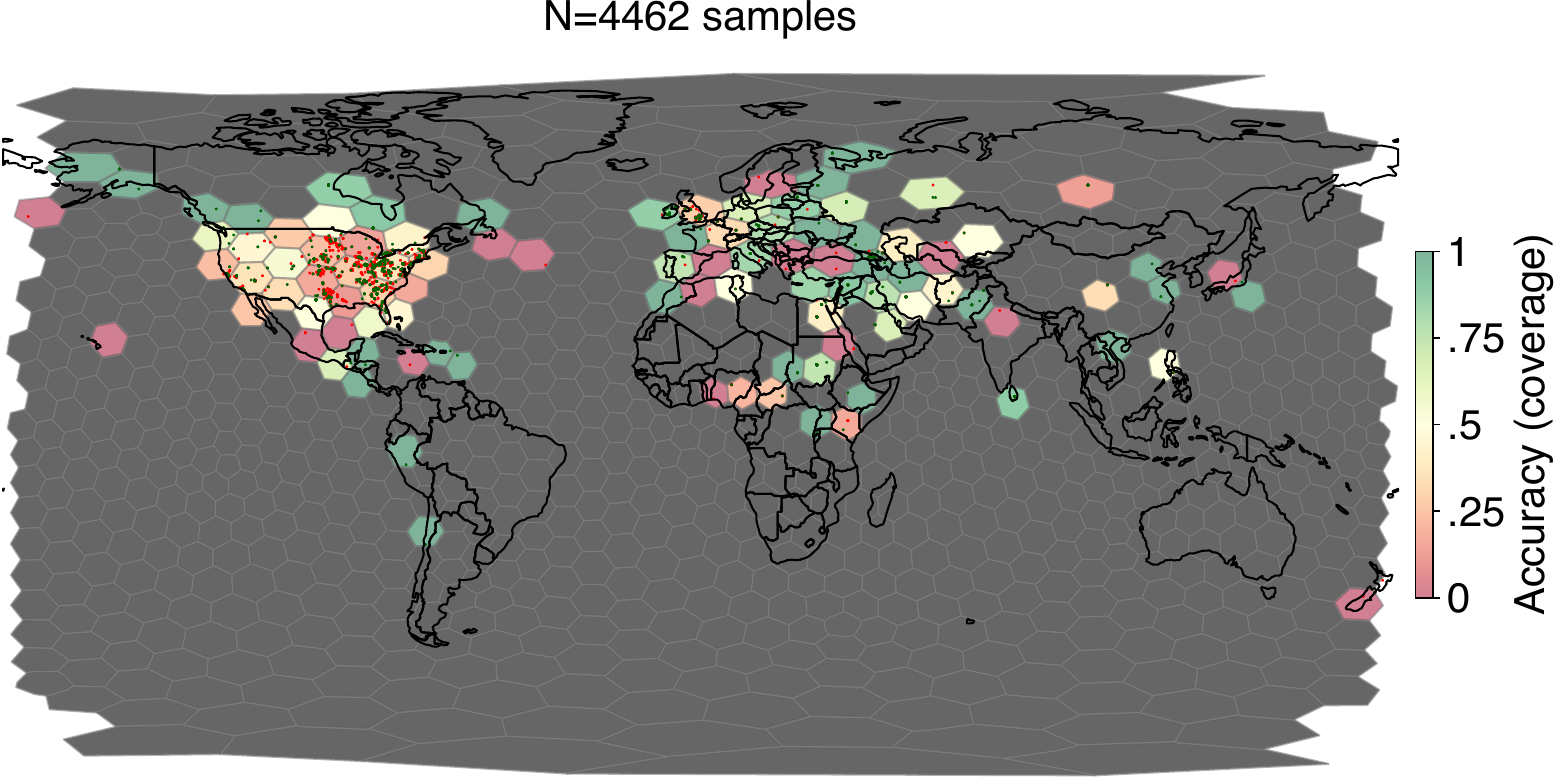}
    }
    \label{fig:map_wiki_acc_ent_link_lgl}
    }
    \hspace{\hshort}
    \subfigure[TR-News]{
    \resizebox*{\imgWidthLeft\columnwidth}{!}{
        \includegraphics{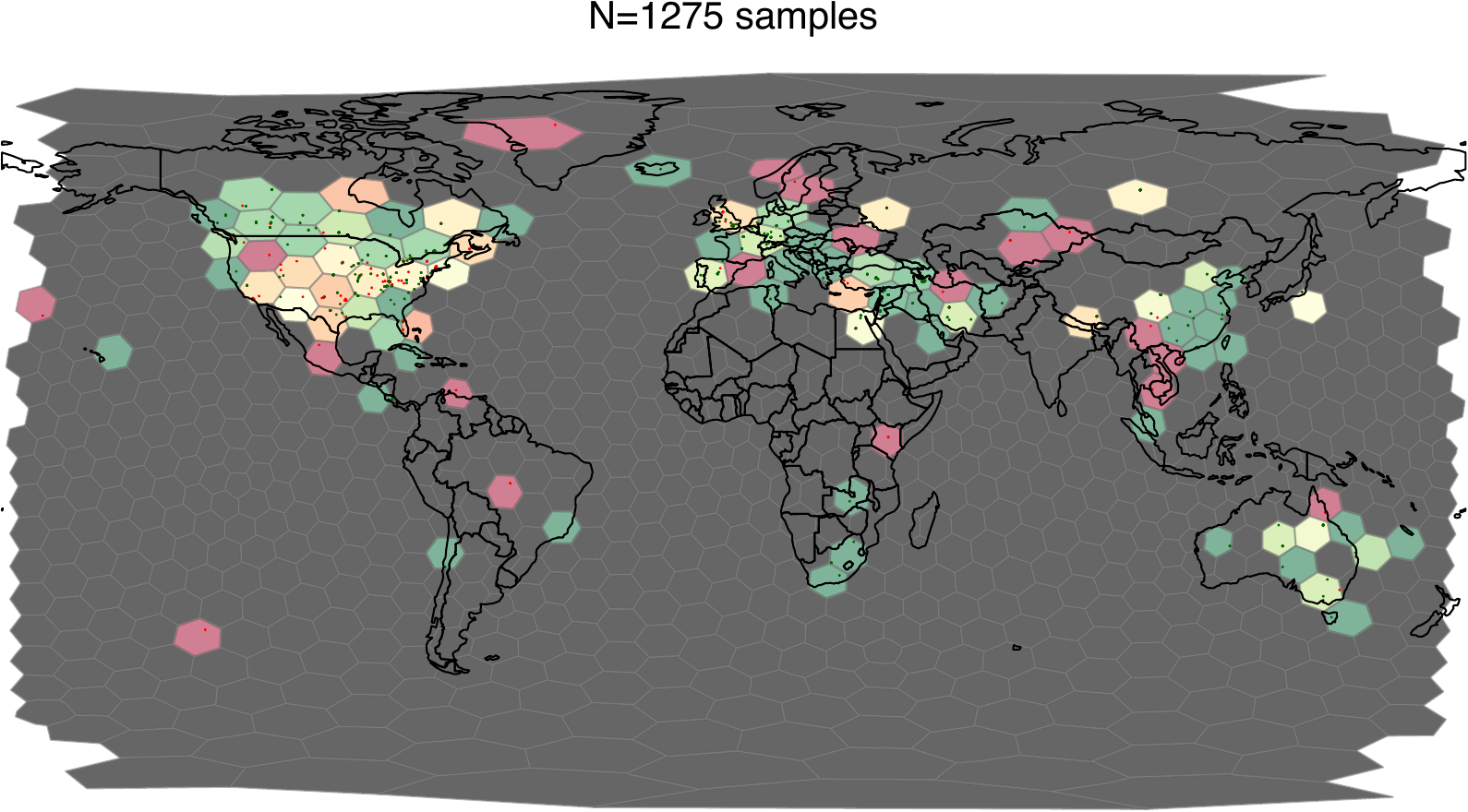}
    }
    \label{fig:map_wiki_acc_ent_link_trnews}
    }   
    \hspace{\hshort}
    \subfigure[WikTor]{
    \resizebox*{\imgWidthRight\columnwidth}{!}{
        \includegraphics{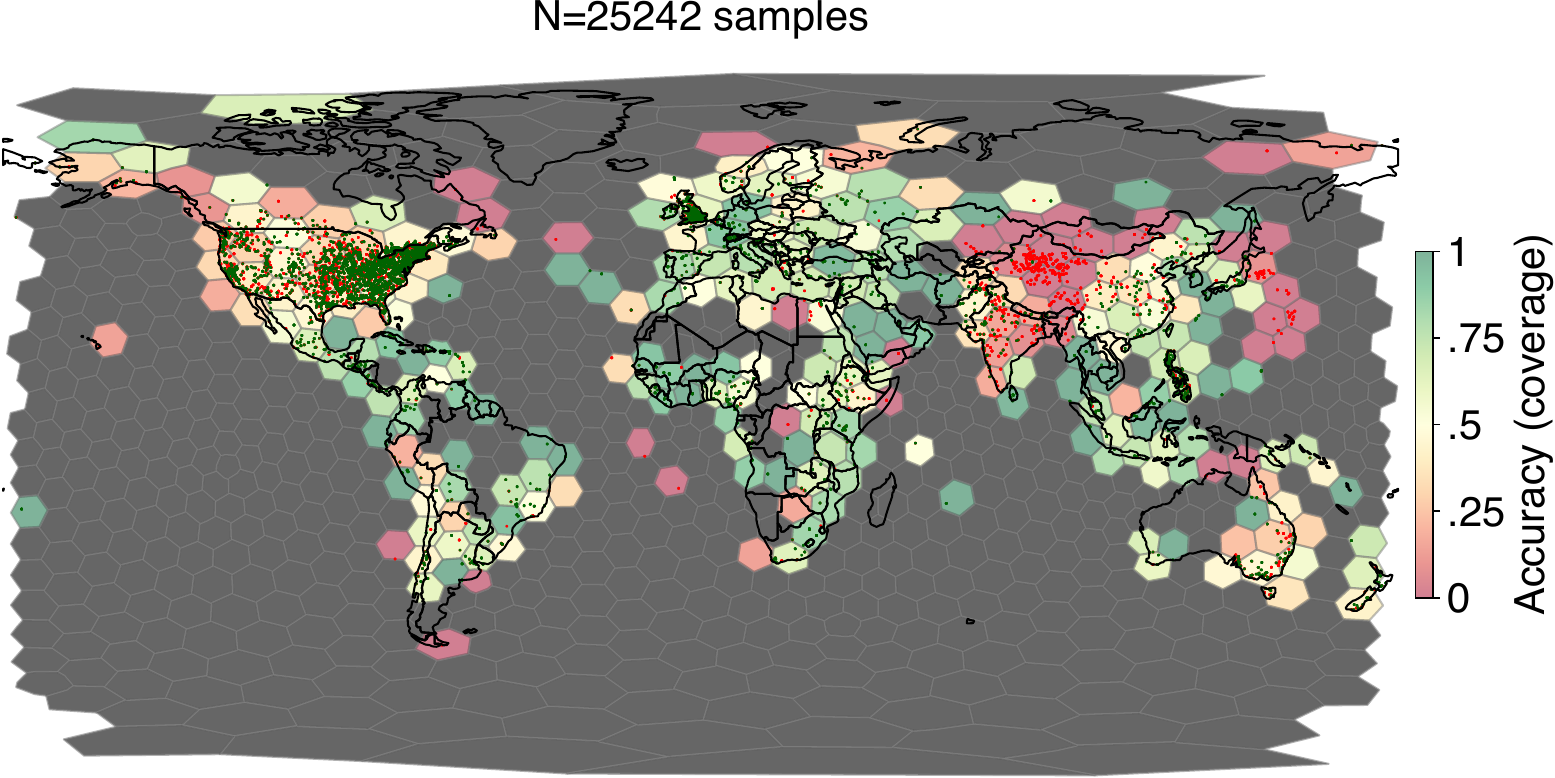}
    }
    \label{fig:map_wiki_acc_ent_link_wiktor}
    }   
    \caption{Local Acc$_{\textrm{coverage}}$ (Gemma with HBS) in the WikiView.} 
    \label{fig:geocoding_mean_err_dist_wiki@1}
\end{figure*}

%% file: 00_CONTENT/09_Discussions.tex
\section{Discussion}

State-of-the-art text encoders demonstrate strong generalization across IR-QA benchmark datasets for \textit{general} knowledge \cite{zhangQwen3EmbeddingAdvancing2025, teamGemma2Improving2024a}. 
Our analysis of zero-shot toponym understanding yields three key insights \textit{specifically} for GeoIR.


\partitle{Text encoders do not generalize well across geographic space.}
Lightweight gazetteer-free architectures have received less attention because 
recent studies report near-perfect toponym resolution in KB-QA setups \cite{huToponymResolutionLeveraging2024}. However, these encoders remain critical performance bottlenecks in modern architectures, such as retrieval-augmented generation systems.

In response, our retrieval framework prioritizes qualitative spatial plausibility over marginal quantitative gains. The proposed methodology values conceptual spatial matches over exact lexical memorization to enforce geographic fidelity. 
Spatial generalization requires models to (i) maintain structural stability, notably during gazetteer pruning with a cut-off resolution; and (ii) adapt toponym recognition for downstream similarity matching.
\textit{Standalone} dense retrieval via text encoders is insufficient for spatial similarity matching here: lexical models (e.g., BM25) remain strong baselines, while text encoders rely primarily on matching surface strings, rather than navigating spatial hierarchies.




\partitle{Gazetteers improve coarse grounding but reveal fine-scale ceilings in text encoders}
Text encoders struggle with direct toponym retrieval. In contrast, generative language models traditionally employ structured reasoning strategies to guide search \cite{yaoReActSynergizingReasoning2022, weiChainofThoughtPromptingElicits2023, yaoTreeThoughtsDeliberate2023}. 
To bridge this gap, our neuro-symbolic search (HBS) introduces a comparable sequential approach for text encoders: the hybrid method constrains the continuous neural representations through a chain of discrete, deterministic geographic steps.

Empirically, combining gazetteers with embeddings improves macro-level accuracy for continents and countries. 
However, inconsistent local representations from text encoders mislead search, leaving retrieval prone to geographically clustered errors that gazetteers cannot resolve.
HBS improves spatial grounding over brute-force search at coarse levels, but chaining retrieval with an explicit spatial index yields limited fine-grained gains. Consequently, embedding models suffer from spatial inaccuracy at fine resolutions, indicating a performance ceiling for neuro-symbolic indexing through administrative hierarchies and frozen vanilla text encoders.

\partitle{English is a low-resource language for space.}
Conceptually, language models \cite{cohenCrawlingInternalKnowledgeBase2023} and gazetteers represent two structurally different knowledge bases: language models rely on repeated contextual associations to learn representations, whereas gazetteers enforce unique spatial identification.

The limited size of human vocabulary restricts expressivity in contextual repetitions, limiting rich toponym representations in training corpora. For comparison, GeoNames indexes 25 million toponyms, whereas an average adult knows fewer than 50,000 lemmas \cite{brysbaertHowManyWords2016} and the English vocabulary contains roughly 1 million words \cite{michelQuantitativeAnalysisCulture2011}. Consequently, language models have shown strong performance on commonsense knowledge, but toponyms represent a specialized vocabulary that lies at the boundary of language model capabilities. Conversely, digital gazetteers \cite{wijegunarathnaDigitalGazetteersReview2025, adamsWahiDiscreteGlobal2017} are also incomplete, capturing a strictly administrative perspective that omits historical records, vernacular names, and multilingual variants.

Alternative perspectives like holonymy \cite{buscaldiConceptualDensitybasedApproach2008}, sense of place \cite{kamruzzamanMakingSensePlace2025}, or discretized geometries  \cite{moltzenLLMAgentsGeorelating2025} also model geographic entities. Text encoders remain \textit{language}-focused and align primarily with semantic views (e.g., WikiView). Designing spatially efficient text encoders requires balancing vocabulary distributions and integrating divergent perspectives---e.g., knowledge graphs---to bridge \textit{administrative} gazetteers and \textit{commonsense} natural language.

%% file: 00_CONTENT/10_Limitations.tex
\section{Limitations}


We identify four main limitations: (1) evaluating frozen encoders measures baseline representations without testing whether gazetteer fine-tuning yields specialized \textit{spatial language} models; (2) space is inherently multi-perspective without a unified metric: point-based distance metrics diverge from coverage-based metrics; (3) skewed spatial hierarchies violate i.i.d. assumptions, complicating statistical significance testing; and (4) gazetteer pruning does not completely eliminate toponym--entry overlap (notably for WikTor in WikiView), risking lexical leakage and score inflation.

%% file: 00_CONTENT/11_Conclusion.tex

\section{Conclusion}

In this paper, we introduced a spatial-semantic indexing framework to \textit{evaluate how language models generalize} across administrative hierarchies rather than relying on exact toponym matching.
Unconstrained embedding retrieval often exhibits limited spatial fidelity. Conversely, incorporating explicit gazetteer structure via hierarchical constraints substantially improves coarse geographic grounding. However, these constraints offer limited benefit---and occasionally degrade performance on---fine-grained point-based metrics; single-scale metrics create artificial discrepancies when evaluating multi-scale geographic entities. 
Developing and evaluating \textit{spatial language} models therefore requires new methodologies that directly account for hierarchical or multi-perspective spatial structures.

These evaluation frameworks are critical as text encoders increasingly serve as primary interfaces in GIS, facing two core challenges in practice.
First, training models solely on distributional semantics leads to mismatches with human spatial reasoning. Second, language models represent a manipulated digital geography subject to intentional alterations (e.g., privacy obfuscation, geo-advertising bias \cite{bourosLocalRec2019Workshop2020}) and malicious threats (e.g., data poisoning \cite{zhaoDataPoisoningDeep}, disputed territory misrepresentation \cite{suiDigitallymediatedTerritorialImaginations2026}). 

Structured gazetteers thus remain essential to calibrate learned representations or evaluate model robustness against geographic incoherence. Nonetheless, human spatial language diverges from normative gazetteers through local subtleties \cite{janowiczWhoseTruthPluralistic2025}. 
Future work will therefore explore how spatial knowledge graphs can bridge this gap by integrating richer semantic perspectives alongside gazetteers, and guiding the development of more grounded \textit{spatial language} models.

%% file: 00_CONTENT/A_Appendix.tex


\section{Toponym ambiguity}
\label{appendix:toponym_ambiguity}

Table~\ref{tab:top_countries_proportion} presents the distribution of ambiguous toponyms across the top-10 most ambiguous countries. In the knowledge graph $\mathcal{K}$, 8,745 of the 51,249 toponyms are homonymous (17.1\%), with these 10 countries accounting for 46.9\% of all ambiguities. The remaining ambiguities mostly involve toponyms shared between these top-10 countries and the rest of the world.

\input{00_CONTENT/FIGS_TABLES_IMG_ALG/stats_kg_ambiguity}

\input{00_CONTENT/FIGS_TABLES_IMG_ALG/algo_hbs}

\section{Brute-force search algorithm}
\label{appendix:algo_brute_force}

The algorithm for Brute-Force Search is presented in Algorithm~\ref{alg:brute_force}.

\section{Hierarchical beam search algorithm}
\label{appendix:algo_hbs}

The algorithm for Hierarchical Beam Search (HBS) is presented in Algorithm~\ref{alg:hierarchical_beam_search}.


\section{Ablation: Enabling early stopping}
\label{appendix:ablation}



We ablate Algorithm~\ref{appendix:algo_hbs} using early stopping: if a candidate toponym $p$ exactly matches mention $t$ (Equation~\ref{def:problem}), search terminates and returns $p$. This acts as a manual early-exit heuristic on easy cases (e.g., countries, provinces) to test whether dense retrieval needs a rule-based shortcut to coarser levels.

Early stopping triggers frequently (Table~\ref{tab:early_stopping_ratio}). Except for BM25, performance gains remain marginal across point-based metrics (Tables~\ref{tab:acc_161_ablation}--\ref{tab:comp_auc_ablation}), 
while coverage-based metric is similar (Table~\ref{tab:acc_coverage_161_ablation}). Hence, standard HBS inherently handles backoff effectively.

\input{00_CONTENT/FIGS_TABLES_IMG_ALG/algo_brute_force}

\input{00_CONTENT/FIGS_TABLES_IMG_ALG/tab_stats_early_stopping}

\input{00_CONTENT/FIGS_TABLES_IMG_ALG/ablation_acc_cov}
\input{00_CONTENT/FIGS_TABLES_IMG_ALG/ablation_acc_161}

\input{00_CONTENT/FIGS_TABLES_IMG_ALG/ablation_med}

\input{00_CONTENT/FIGS_TABLES_IMG_ALG/ablation_auc}






%% file: 00_CONTENT/FIGS_TABLES_IMG_ALG/stats_kg_ambiguity.tex

\begin{table}[htbp]
\centering
\Small
\caption{Top-10 countries with highest proportion of toponyms. \textbf{Global Share} indicates the country's toponym count over total global dataset toponyms. \textbf{Internal Ratio} indicates the number of internally homonymous toponyms over the total toponyms within that country.}
\label{tab:top_countries_proportion}
\begin{tabular}{lll}
\toprule
\textbf{Country} & \textbf{Global Share} & \textbf{Internal Ratio} \\
\midrule
BRA & 5,600 / 51,249 {\small (10.9\%)} & 523 / 5,600 {\small (9.3\%)} \\
USA & 3,200 / 51,249 {\small (6.2\%)} & 1,779 / 3,200 {\small(55.6\%)} \\
ROU & 2,982 / 51,249 {\small (5.8\%)} & 573 / 2,982 {\small (19.2\%)} \\
RUS & 2,529 / 51,249 {\small (4.9\%)} & 314 / 2,529 {\small (12.4\%)} \\
MEX & 2,490 / 51,249 {\small (4.9\%)} & 260 / 2,490 {\small (10.4\%)} \\
JPN & 1,859 / 51,249 {\small (3.6\%)} & 308 / 1,859 {\small (16.6\%)} \\
PHL & 1,729 / 51,249 {\small (3.4\%)} & 359 / 1,729 {\small (20.8\%)} \\
DZA & 1,553 / 51,249 {\small (3.0\%)} & 160 / 1,553 {\small (10.3\%)} \\
COL & 1,153 / 51,249 {\small (2.2\%)} & 172 / 1,153 {\small (14.9\%)} \\
TUR & 1,011 / 51,249 {\small (2.0\%)} & 114 / 1,011 {\small (11.3\%)} \\
\bottomrule
\end{tabular}
\end{table}

%% file: 00_CONTENT/FIGS_TABLES_IMG_ALG/algo_hbs.tex
\noindent\begin{minipage}{\linewidth}
\small
\noindent\hrulefill
\vspace{-2.7ex}
\captionsetup{
    justification=raggedright, 
    singlelinecheck=false, 
    labelfont=bf,           
    labelsep=space,         
    textfont=normalfont     
}
\captionof{algorithm}{Hierarchical Beam Search (HBS)}
\label{alg:hierarchical_beam_search}
\vspace{-1.7ex}
\noindent\hrulefill
\begin{algorithmic}[hbt]
\renewcommand{\algorithmicrequire}{\textbf{Inputs:}}
\Require 
\Statex $\triangleright$ Dataset $\mathcal{D}$ containing texts $[t, T]$, with $t$ a toponym, and $T$ the surrounding context (window = 50);
\Statex $\triangleright$ Query template $Q$ with place-holders for $t$ and $T$;
\Statex $\triangleright$ Knowledge graph $\mathcal{K}$ with $p$ toponyms of reference, associated with a text view $D_{p, \texttt{view}}$;
\Statex $\triangleright$ Hierarchical levels $H=$  (Continent, Country, Province, City);
\Statex $\triangleright$ Beam values list $\mathcal{B} = (b_h, \dots)$ for level $h \in H$;
\Statex $\triangleright$ Maximum number $n_\textrm{max}$ of candidates $p$ in each retrieval step.

\State \textcolor{gray}{\% Pre-compute embeddings for each toponym of reference}
\For{each $p \in \mathcal{K}$}
    \State $\mathbf{e}_p \gets \texttt{Enc}(D_{p, \texttt{view}})$
\EndFor

\State \textcolor{gray}{\% Process each toponym mention in the dataset}
\For{each $[t, T] \in \mathcal{D}$}
    \State \textcolor{gray}{\% Populate the query template and encode}
    \State $q_{\text{input}} \gets Q([t, T])$
    \State $\mathbf{q} \gets \texttt{Enc}(q_{\text{input}})$
    
    \State \textcolor{gray}{\% Initialization / Pre-selection}
    \State $\mathcal{C}_{\text{valid}} \gets \{ p \in \mathcal{K} \mid (p, \text{has\_level}, \text{continent}) \}$
    
    \State \textcolor{gray}{\% Iterate through spatial hierarchy levels}
    \For{each level $h \in H$}
        \State \textcolor{gray}{\% Filter candidates by parent hierarchy}
        \If{$h = \text{Continent}$}
            \State $\mathcal{P}_{\text{candidates}} \gets \mathcal{C}_{\text{valid}}$
        \Else
            \State $\mathcal{P}_{\text{candidates}} \gets \{ p \in \mathcal{K} \mid \texttt{parent\_of}(p) \in \mathcal{C}_{\text{valid}} \}$
        \EndIf
        
        \State \textcolor{gray}{\% Select top-$b_h$ candidates based on semantic similarity}
        \State $\mathcal{C}_{\text{valid}} \gets \texttt{top-b}_\texttt{h} \{ \mathrm{sim}(\mathbf{q}, \mathbf{e}_p) \mid p \in \mathcal{P}_{\text{candidates}} \}$
        

        \State \textcolor{gray}{\% Prune the results for memory}
        \State $\mathcal{C}_{\text{valid}} \gets \texttt{top-n}_\texttt{max} \left( \mathcal{C}_{\text{valid}} \right)$
        
    \EndFor
    
    \State \Return $\mathcal{C}_{\text{valid}}$
\EndFor
\end{algorithmic}
\vspace{-1em}
\noindent\hrulefill
\vspace{1em}
\end{minipage}


%% file: 00_CONTENT/FIGS_TABLES_IMG_ALG/algo_brute_force.tex
\begin{algorithm}[H]
\small
\caption{Brute-Force Search}
\label{alg:brute_force}
\begin{algorithmic}[1]
\renewcommand{\algorithmicrequire}{\textbf{Inputs:}}
\Require 
\Statex $\triangleright$ Dataset $\mathcal{D}$ containing texts $[t, T]$, with $t$ a toponym, and $T$ the surrounding context (window = 50);
\Statex $\triangleright$ Query template $Q$ with place-holders for $t$ and $T$;
\Statex $\triangleright$ Knowledge graph $\mathcal{K}$ with $p$ toponyms of reference, associated with a text view $D_{p, \texttt{view}}$;
\Statex $\triangleright$ Maximum number $n_\textrm{max}$ of candidates $p$ in the final step.

\State \textcolor{gray}{\% Pre-compute embeddings for each toponym of reference}
\For{each $p \in \mathcal{K}$}
    \State $\mathbf{e}_p \gets \texttt{Enc}(D_{p, \texttt{view}})$
\EndFor

\State \textcolor{gray}{\% Process each toponym mention in the dataset}
\For{each $[t, T] \in \mathcal{D}$}
    \State \textcolor{gray}{\% Populate the query template and encode}
    \State $q_{\text{input}} \gets Q([t, T])$
    \State $\mathbf{q} \gets \texttt{Enc}(q_{\text{input}})$
    
    
        
        \State \textcolor{gray}{\% Select top-$n_\textrm{max}$ candidates based on semantic similarity}
        \State $\mathcal{C}_{\text{valid}} \gets \texttt{top-n}_\texttt{max} \{ \mathrm{sim}(\mathbf{q}, \mathbf{e}_p) \mid p \in \mathcal{K} \}$
        
        \State \textcolor{gray}{\% Check if exact text match is found and end}
        \For{each $p \in \mathcal{C}_{\text{valid}}$}
            \If{$\texttt{text}(p) = \texttt{text}(t)$}
                \State \Return $p$ \text{and complete with the other top $n_\texttt{max} - 1$}
            \EndIf
        \EndFor

        
    \EndFor

    \State \textcolor{gray}{\% Return the retrieved set by default}
    \State \Return $\mathcal{C}_{\text{valid}}$
\end{algorithmic}
\end{algorithm}

%% file: 00_CONTENT/FIGS_TABLES_IMG_ALG/tab_stats_early_stopping.tex
\begin{table}[htbp]
\Small
\centering
\caption{Early stopping frequency.}
\label{tab:early_stopping_ratio}
\begin{NiceTabular}{l c c c c | c c c c}
\toprule
& \multicolumn{4}{c}{GeoView} & \multicolumn{4}{c}{WikiView} \\
Model & GV & LGL & TRN & WT & GV & LGL & TRN & WT \\
\midrule
BM25 & .57 & .17 & .27 & .25 & .61 & .20 & .36 & .28\\
MiniLMv2 & .50 & .17 & .30 & .18 & .63 & .25 & .44 & .27\\
ColBERTv2 & .34 & .10 & .18 & .11 & .36 & .13 & .25 & .13\\
BGE base & .64 & .36 & .47 & .33 & .67 & .41 & .50 & .43\\
Qwen3 & .54 & .16 & .35 & .20 & .67 & .32 & .50 & .34\\
Gemma & .68 & .29 & .48 & .34 & .67 & .32 & .49 & .35\\
\bottomrule
\end{NiceTabular}
\end{table}

%% file: 00_CONTENT/FIGS_TABLES_IMG_ALG/ablation_acc_cov.tex
\begin{table}[H]
  \caption{Comparison for Acc$_{\text{coverage}}$ ($\uparrow$). Best for the configuration: {\large .91}; best for the dataset: \underline{.91}.}
  \label{tab:comp_0}
  \centering
  {\Small
  \begin{tabular}{llcccc|cccc}
    \toprule
    & & \multicolumn{4}{c}{HBS (w/o early stopping)} & \multicolumn{4}{c}{HBS (w/ early stopping)} \\
    View & Model & GV & LGL & TRN & WT & GV & LGL & TRN & WT\\
    \midrule
    \multirow{6}{*}{Geo} & BM25 & .69 & .24 & .40 & .49 & .69 & .23 & .38 & .48 \\
    & MiniLMv2 & .65 & .31 & .48 & .33  & .66 & .30 & .49 & .32\\
    & ColBERTv2 & .52 & .21 & .39 & .30 & .48 & .16 & .34 & .27\\
    & BGE base & .84 & \textbf{.71} & .74 & .64  & .82 & .70 & .74 & .63\\
    & Qwen3  & .67 & .24 & .50 & .32  & .68 & .23 & .51 & .32\\
    & Gemma & \textbf{.90} & .56 & \textbf{.76} & \textbf{.68}  & .89 & .55 & .74 & \textbf{.68}\\
    \midrule
    \multirow{6}{*}{Wiki} & BM25 & .72 & .29 & .51 & .53  & .78 & .30 & .53 & .51 \\
    & MiniLMv2 & .77 & .40 & .63 & .44  & .76 & .39 & .63 & .44 \\
    & ColBERTv2 & .47 & .18 & .40 & .25  & .48 & .17 & .40 & .21 \\
    & BGE base & \best{.91} & \best{.87} & \best{.85} & \best{.81}  & .88 & .84 & .84 & .80 \\
    & Qwen3 & .88 & .54 & .75 & .60  & .86 & .52 & .75 & .60 \\
    & Gemma & .86 & .55 & .76 & .69  & .88 & .54 & .77 & .66 \\
    \bottomrule
  \end{tabular}
  }
  \label{tab:acc_coverage_161_ablation}
\end{table}

%% file: 00_CONTENT/FIGS_TABLES_IMG_ALG/ablation_acc_161.tex
\begin{table}[hbt]
  \caption{Comparison for Acc@161km ($\uparrow$). Best for the experiment noted as {\large .56}; and best for the dataset as \underline{.56}.}
  \label{tab:comp_0}
  \centering
  {\Small
  \begin{NiceTabular}{llcccc|cccc}
    \CodeBefore
      \tikz \fill [pattern=north east lines, pattern color=gray!80] (3-|7) rectangle (5-|11);
    \Body
    \toprule
    & & \multicolumn{4}{c}{HBS (w/o early stopping)} & \multicolumn{4}{c}{HBS (w/ early stopping)} \\
    View & Model & GV & LGL & TRN & WT & GV & LGL & TRN & WT\\
    \midrule
    & \opt{CamCoder}  & \textbf{\opt{.82}} & \textbf{\opt{.76}} & \textbf{\opt{.67}} & \opt{.65}  &  &  &  &  \\ 
    & \opt{MLG}  & \opt{.49} & \opt{.53} & \opt{-} & \textbf{\opt{.91}}  &  &  &  &  \\
    \midrule
    \multirow{6}{*}{Geo} & BM25 & .13 & .05 & .07 & .22 & .41 & .14 & .19 & .27 \\
    & MiniLMv2 & .36 & .17 & .24 & .15 & .36 & .16 & .24 & .15 \\
    & ColBERTv2 & .24 & .09 & .14 & .11 & .24 & .09 & .14 & .11 \\
    & BGE base & .47 & .35 & .37 & .34 & .46 & .35 & .37 & .33 \\
    & Qwen3 & .39 & .16 & .26 & .19  & .39 & .14 & .25 & .19 \\
    & Gemma & \best{.56} & \textbf{.36} & \textbf{.45} & \textbf{.43} & .53 & .34 & .41 & \textbf{.43}\\
    \midrule
    \multirow{6}{*}{Wiki} & BM25 & .29 & .15 & .23 & .33 & .46 & .19 & .31 & .29 \\
    & MiniLMv2 & .44 & .24 & .35 & .22 & .44 & .23 & .35 & .22 \\
    & ColBERTv2 & .23 & .10 & .15 & .10 & .23 & .09 & .15 & .09 \\
    & BGE base & \textbf{.53} & \best{.45} & \best{.47} & .43 & .49 & .41 & .43 & \best{.44} \\
    & Qwen3 & \textbf{.53} & .34 & .42 & .31 & .50 & .31 & .41 & .31 \\
    & Gemma & .46 & .35 & .44 & \best{.44}  & .52 & .35 & .46 & .40 \\
    \bottomrule
    \end{NiceTabular}
  \label{tab:acc_161_ablation}
  }
\end{table}

%% file: 00_CONTENT/FIGS_TABLES_IMG_ALG/ablation_med.tex
\begin{table}[hbt]
  \caption{Comparison of MED ($\downarrow$). 7{,}000~km roughly represents the distance between the USA and Europe; 4{,}000~km between the west and east coasts of the USA; 1{,}000~km between the Western and Eastern coasts of Spain. Best for the experiment noted as {\large .10}; and best for the dataset as \underline{.10}.}
  \label{tab:comp_med_ablation}
  \centering
  {\Small
  \begin{NiceTabular}{llcccc|cccc}
    \CodeBefore
      \tikz \fill [pattern=north east lines, pattern color=gray!80] (3-|7) rectangle (5-|11);
    \Body
    \toprule
    & & \multicolumn{4}{c}{HBS (w/o early stopping)} & \multicolumn{4}{c}{HBS (w/ early stopping)} \\
    View & Model & GV & LGL & TRN & WT & GV & LGL & TRN & WT \\
    \midrule
    & \opt{CamCoder}  & \textbf{\opt{7}} & \opt{22} & \textbf{\opt{30}} & \opt{11}  &  &  &  &  \\ 
    & \opt{MLG} & \opt{17} & \textbf{\opt{14}} & \opt{-} & \textbf{\opt{2}}  &  &  &  &  \\
    \midrule
    \multirow{6}{*}{Geo} & BM25 & 31 & 59 & 53 & 48 & 27 & 54 & 51 & 48 \\
    & MiniLMv2 & 28 & 61 & 42 & 57 & 28  & 65 & 43 & 61 \\
    & ColBERTv2 & 37 & 63 & 47 & 52 & 39  & 69 & 53 & 55 \\
    & BGE base & 14 & \textbf{25} & 22 & 30 & 14  & \textbf{25} & 21 & 30 \\
    & Qwen3 & 25 & 66 & 41 & 52 & 26  & 71 & 44 & 56\\
    & Gemma & \textbf{11} & 38 & \textbf{19} & \textbf{27} & \textbf{11} & 39 & 21 & \textbf{27} \\
    \midrule
    \multirow{6}{*}{Wiki} & BM25 & 27 & 66 & 46 & 44 & 22 & 68 & 49 & 46 \\
    & MiniLMv2 & 20 & 51 & 32 & 47 & 20 & 52 & 33 & 48 \\
    & ColBERTv2 & 39 & 59 & 48 & 53 & 35 & 64 & 49 & 60 \\
    & BGE base & \best{10} & \best{16} & \best{16} & \best{20} & 12 & 17 & \best{16} & \best{20} \\
    & Qwen3 & 11 & 38 & 22 & 34  & 12 & 41 & 24 & 35 \\
    & Gemma & 13 & 39 & 21 & 27 & 12 & 39 & 21 & 29 \\
    \bottomrule
    \end{NiceTabular}
  }
\end{table}

%% file: 00_CONTENT/FIGS_TABLES_IMG_ALG/ablation_auc.tex
\begin{table}[hbt]
  \caption{Comparison for AUC ($\downarrow$). Best for the experiment: {\large .52}; for the dataset: \underline{.52}.}
  \label{tab:comp_auc_ablation}
  \centering
  {\Small
  \begin{NiceTabular}{llcccc|cccc}
    \CodeBefore
      \tikz \fill [pattern=north east lines, pattern color=gray!80] (3-|7) rectangle (5-|11);
    \Body
    \toprule
    & & \multicolumn{4}{c}{HBS (w/o early stopping)} & \multicolumn{4}{c}{HBS (w/ early stopping)} \\
    View & Model & GV & LGL & TRN & WT & GV & LGL & TRN & WT \\
    \midrule
    & \opt{CamCoder}  & \opt{.65} & \opt{.60} & \textbf{\opt{.31}} & \opt{.49}  &  &  &  &  \\ 
    & \opt{MLG}  & \textbf{\opt{.55}} & \textbf{\opt{.54}} & \opt{-} & \textbf{\opt{.37}}  &  &  &  &  \\
    \midrule
    \multirow{6}{*}{Geo} & BM25 & .71 & .82 & .79 & .73 & .61 & .79 & .75 & .72 \\
    & MiniLMv2 & .63 & .78 & .71 & .77 & .63  & .79 & .71 & .78 \\
    & ColBERTv2 & .70 & .82 & .76 & .78 & .70  & .83 & .77 & .79 \\
    & BGE base & .56 & \textbf{.62} & .60 & .64 & .56  & .63 & .61 & .64 \\
    & Qwen3 & .61 & .80 & .69 & .75 & .61  & .81 & .70 & .76 \\
    & Gemma & \best{.52} & .66 & \textbf{.57} & \textbf{.60} & \best{.52}  & .67 & .59 & \textbf{.60} \\
    \midrule
    \multirow{6}{*}{Wiki} & BM25 & .64 & .80 & .72 & .68 & .58 & .79 & .68 & .69 \\
    & MiniLMv2 & .58 & .73 & .65 & .72 & .58 & .74 & .65 & .72 \\
    & ColBERTv2 & .70 & .81 & .75 & .79 & .69 & .82 & .75 & .80 \\
    & BGE base & \textbf{.53} & \best{.57} & \best{.56} & \best{.57} & .54 & .59 & .58 & \best{.57} \\
    & Qwen3 & \textbf{.53} & .67 & .59 & .65 & .54 & .68 & \best{.56} & .66 \\
    & Gemma & .55 & .67 & .58 & .59 & \textbf{.53} & .67 & .57 & .62 \\
    \bottomrule
  \end{NiceTabular}
  }
\end{table}